\documentclass{IEEEtran}
\usepackage{upgreek}
\usepackage{combelow}
\usepackage{comment}
\usepackage{amsmath,amssymb,amsfonts}
\usepackage{algorithmic}
\usepackage{textcomp}
\def\BibTeX{{\rm B\kern-.05em{\sc i\kern-.025em b}\kern-.08em
    T\kern-.1667em\lower.7ex\hbox{E}\kern-.125emX}}
\usepackage{tikz,graphicx}
\usepackage{color}
\usepackage[mathscr]{eucal}

\def\XXint#1#2#3{{\setbox0=\hbox{$#1{#2#3}{\int}$}
     \vcenter{\hbox{$#2#3$}}\kern-.5\wd0}}

\newcommand{\be}[1]{\begin{equation} \label{#1} }
\newcommand{\bea}[1]{\begin{eqnarray} \label{#1} }

\newcommand{\bfi}{\begin{figure}}
\newcommand{\efi}{\end{figure}} 
\newcommand{\ee}{\end{equation}}
\newcommand{\eea}{\end{eqnarray}}
\newcommand{\bib}{\bibitem}
\newcommand{\lbl}{\label}

\newcommand{\w}{{\omega}}

\newcommand{\vr}{{\bf r}}

\newcommand{\vE}{{\bf E}}

\newcommand{\vrho}{{\mbox{\boldmath $\rho$}}}

\newcommand{\la}{\lesssim}
\newcommand{\ga}{\gtrsim}
\newcommand{\eps}{\epsilon}

\begin{document}
\title{\LARGE\bf Stellar Speckle and Correlation Derived from Classical Wave Expansions for Spherical Antennas}
\author{Arthur D. Yaghjian\\\textit{\normalsize Electromagnetics Research, Concord, MA  01742,  USA} ({\small a.yaghjian@comcast.net}) 
}
\maketitle
\begin{abstract}
Michelson phase and Hanbury Brown--Twiss intensity stellar interferometry require expressions for the first- and second-order correlation functions, respectively, of the fields radiated by stars in terms of their diameters and measured quasi-monochromatic wavelengths.  Although our sun and most other stars are spherical in shape at optical wavelengths, previous determinations of speckle and correlation functions have modeled stars as circular discs rather than spheres because of the mathematical tools available for partially coherent fields on planar surfaces.  However, with the incentive that most stars are indeed shaped like spheres and not discs, the present paper models a star as a spherical antenna composed of a random distribution of uncorrelated volume sources within a thin surface layer (photosphere).  Working directly with the time-domain fields, a self-contained, straightforward, detailed derivation of speckle patterns and correlation functions is given based on a novel, angularly symmetric, spherical mode expansion with coefficients determined by the assumed Lambertian nature of the star's radiation and the uniform asymptotic behavior of the spherical Hankel functions. First-order spatially averaged and temporally averaged correlation functions are proven to be identical and the normalized second-order correlation function is shown to equal one plus the square of the normalized first-order correlation function.  The direct time-domain approach reveals explicit expressions for the quasi-monochromatic wave-packet fields of stellar radiation as well as new criteria for the validity of the far-field approximation for the fields of incoherent sources that are much less restrictive than the Rayleigh-distance criterion for coherent sources.
\end{abstract}
\section{Introduction}\label{sec:Introduction}
\IEEEPARstart{H}{aving} studied and worked within the area of electromagnetics for about 50 years, it bothered me that I (and apparently most of my colleagues) understood so little about the properties of sunlight, the electromagnetic fields most responsible for life on Earth.  Most of our education and research in electromagnetics dealt mainly with coherent sources of radiation, that is, single-frequency (monochromatic) sources with specified or determined positions, magnitudes and phases.  Whereas radiation from the sun and other stars is produced by incoherent sources, that is, randomly positioned sources emitting a spectrum of frequencies that for narrow-bandwidth reception produce quasi-monochromatic wave packets rather than continuous-wave fields.  It is generally known that these wave packets create a ``speckle pattern'' on Earth that rapidly changes with time and that the retinal rods and cones of our eyes,  photographic film, or photoelectric detectors record an average intensity at different wavelengths that are chemically resolved by the cones of the eye and the silver halide layers of color film, or by filters of photoelectric detectors.  For a deeper understanding of fields radiated by stars, one can consult the relatively few textbooks in electromagnetics and optics that contain the basic theory of incoherent sources, for example, \cite{B&W}--\cite{Rastogi}.  However, these texts concentrate on the radiation or scattering from planar surfaces and approximate the sun and other stars within the visible spectrum by their projected circular discs rather than by spheres.   Also, because a primary aim of most of these texts is to develop a general theory of coherence, their treatment of stellar radiation involves the introduction of complex analytic signals and ensemble averages, as well as the deriving or invoking of the van Cittert-Zernicke and central limit theorems, and the moment theorem for Gaussian random processes.  This general statistical theory is important for those who plan to continue work in the area of partial coherence.  However, it is a formidable amount of background material to master for someone who wants only to understand stellar radiation.
\par
There are definite reasons that the sun and other stars have historically been modeled by circular discs for the sake of determining speckle and interferometric correlation functions even though stars are spherical in shape. The fields radiated by a star's projected disc are determined by the integral of the free-space Green's function multiplied by either the tangential electric or magnetic field over the planar aperture of the disc.\footnote{The limits of integration cover the entire infinite plane of the disc but the tangential field outside the projected disc area is usually presupposed (without proof) to contribute negligibly to the far fields in the directions close to the normal to the disc.}   In contrast to the disc, as Beran and Parrent point out \cite[p. 67]{B&P},\footnote{The 1964 statement of Beran and Parrent that the ``solution for [partially coherent] radiation from other than plane surfaces is extremely difficult, and little has been done in this area" still applies today with the exception of the paper by Agarwal, Gbur, and Wolf \cite{AGW}.  This brief paper concentrates on the spherical wave expansion for the cross spectral density and makes the simplifying assumption that the propagating fields near the surface of the sun are fully incoherent (exhibit delta-function correlation).  They do not obtain expressions for the fields or their correlation functions but numerically evaluate the spherical wave expansion for the cross spectral density.}  a closed-form expression for the Dirichlet or Neumann Green's function for the sphere does not exist and, thus, both the tangential electric and magnetic fields are required in a tractable expression for the fields radiated by a sphere in terms of an integral of the fields over the surface of the sphere.  Moreover, for the sun or other stars, one does not know the relationship between the electric and magnetic fields at each point on the surface circumscribing the sun or another star.
\par
\begin{figure}[!t]
\centerline{\includegraphics[width=5cm]{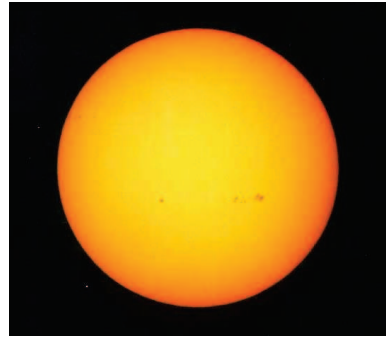}}
\vspace{-1mm}
\caption{Color photograph of the sun showing limb darkening and reddening.}
\label{figredsun}
\vspace{-3mm}
\end{figure}

Another reason that the disc model of the sun and other stars proves convenient is that it allows one to readily incorporate an observed intensity taper (``limb effects'') toward the circular edge of the sun or other stars into the aperture fields of the circular planar disc.\footnote{The intensity taper of the sun (or another star) observed with the eye or a telescope is called \textit{limb darkening} (or \textit{brightening} if the star appears brighter toward its edge). Limb darkening of the sun occurs because most of the light comes from a fixed ``optical depth'' in the photosphere along the line of sight, and the temperature (and thus the light intensity) increases with depth into the photosphere.  When we look toward the edge (limb -- from the Latin ``limbus'') of the sun, we see the light that comes through a shallower portion of the photosphere with a somewhat lower temperature.  The same lower temperature explains why the sun's light is redder toward its limb; see Fig. \ref{figredsun}.}
However, it can be proven for planar surfaces (like the projected disc of a star) that the surface fields, unlike point volume sources, cannot be spatially fully uncorrelated (incoherent) but are approximately sinc-function correlated with the first null at about a half-wavelength separation distance \cite[sec. 5.5]{Wolf}.\footnote{According to the Whittaker-Shannon sampling theorem \cite[sec. 2.4.1]{Goodman}, narrow-band propagating fields just outside the surface of the star are completely determined by their values separated by about half a mean wavelength.  Thus, it follows that the propagating tangential fields are always correlated for separation distances less than about half a mean wavelength.}  This implies that it is not strictly valid to use delta correlation functions (zero correlation for separation distances greater than zero) for the surface fields of the disc, as is commonly done, to obtain the classic far-field stellar zero-correlation angles ($1.22\lambda_0/D$ for a uniform-intensity disc, where $\lambda_0$ is the mean wavelength and $D=2a$ is the diameter of the disc/star).
\par
Of course, one can simply ignore the fact that the projected disc of the sun or star lies in free space and assume that the disc is composed of a thin layer of uncorrelated volume sources \cite[sec. 3.2]{Wolf}.  However, this unrealistic assumption is unnecessary because, in directions of observation close to the normal to the disc, sinc-like half-wavelength or so surface-field correlation distances produce practically the same far-field angular correlation as the delta surface-field correlation function \cite[pp. 63-64, 67]{B&P}.\footnote{The practical differences between sinc-correlated and delta-correlated surface fields become significant only for angles $\alpha$ from the normal to the disc such that $\cos\alpha$ is appreciably less than unity because for half-wavelength sampling of the fields, the tangential surface field is practically uncorrelated and can be replaced by uncorrelated volume sources at the same sample points since the $\cos\alpha$ factor multiplying the single tangential-field diffraction integral close to the normal is approximately equal to unity, like that of the volume-source diffraction integral, which has no $\cos\alpha$ factor.}  Still, this justification of the delta surface-field correlation function for observation directions close to the normal to the disc does not render the disc model for a spherical sun or other star entirely satisfactory since it is unclear how to rigorously determine the actual surface-field correlation function throughout the projected planar disc in front of the sun or other star and, finally, the sun and most other stars are shaped like spheres not discs.
\par
Consequently, the primary purpose of the present paper is to consider the sun and other stars as spherical antennas composed of a random distribution of uncorrelated volume sources radiating from a thin surface atmosphere in a visible bandwidth that is narrow (quasi-monochromatic) but not single-frequency (monochromatic).  Working directly with the real time-domain fields and without explicit use or derivation of the van Cittert-Zernicke, central limit, or moment theorems, first- and second-order correlation functions, which determine average angular speckle width in terms of star diameter and mean wavelength, are derived and related to each other for Michelson phase stellar interferometry and for Hanbury Brown--Twiss intensity stellar interferometry, respectively.  In addition, the first-order spatially and temporally averaged correlation functions are shown to be identical and the second-order normalized  correlation function is shown to equal the square of the first-order normalized correlation function plus one.
\par
Although we neglect limb darkening/brightening in the analysis of this paper, as well as granulation, brightspots (sunspots), and faculae,  so that the sun and other stars are assumed to be ``Lambertian radiators'', the effect of limb darkening/brightening can readily be incorporated into the analysis --- specifically, by inserting the limb darkening/brightening (intensity taper) function into the far-field intensity factor, as explained in Section \ref{subsec:Lambertian}.
\par
The derivation is made possible by expanding the fields of the star in a finite number of newly derived, angularly sym\-me\-tric spherical modes whose coefficients have statistically in\-de\-pen\-dent phase and an average magnitude determined by combining the Lambertian radiation of the star with the uniform asymptotic expansion of the spherical Hankel func\-tions.  In this way, we avoid the difficulties in relating the electric and magnetic near fields over the surface of the circumscribing sphere and in deciding how these surface fields are correlated.  Although the correlation functions for the spherical model of a star do not differ appreciably from those obtained previously near the normal directions of a uniform-intensity stellar disc model (the disc uses a $\sin x\! \approx\! x$ approximation), they are derived from the real time-domain fields using a relatively simple, straightforward, self-contained antenna analysis applied to a more realistic, less restrictive spherical stellar model.  In addition, the straightforward time-domain analysis reveals explicit expressions for the quasi-monochromatic received fields radiated by the stars (not just the cor\-re\-la\-tion functions), and new formulas (much less restrictive than the coherent-field Rayleigh distance) for the radial distance from the star beyond which the partially coherent quasi-monochromatic fields are given by their far-field approximations. 
\pagenumbering{arabic}
\setcounter{page}{2}
\subsection{Idealized Spherical Model of a Star}\label{subsec:Idealized}
We will assume a model of the star in which the predominant radiation for each wavelength of visible light is produced by volume sources within a thin surface atmosphere (photosphere) with an outer radius $a$.   Inside the thin surface atmosphere (thickness equal to a small fraction of $a$), there is a high density of statistically independent (uncorrelated) volume-source radiators (mainly hydrogen and helium molecules for the sun) distributed with spherical symmetry (on a macroscopic level).  The outer radius $a$ of the thin surface atmosphere of the sun or similar star is assumed to lie in free space just outside the significant reactive (evanescent) fields of the sources, that is, there is free space without significant reactive fields in the infinite volume $r \ge a$, where the radial distance $r$ is measured from the center of the sphere.  Moreover, it is assumed that the radius $a$ of the star is the same for each wavelength in the observed bandwidth of the visible spectrum or, equivalently, that the change in the radius over the observed bandwidth is a negligible fraction of the mean radius designated by $a$.  In addition, we assume negligible limb effects, that is, the star is a Lambertian radiator.
\par
Of course, these assumptions do not hold perfectly.  The observed outer radius of the sun changes measurably over appreciable bandwidths.  The variation in temperature across the finite thickness of the surface atmosphere can produce limb darkening or brightening, depending on the frequency, yet our spherical model neglects limb effects.  For the purpose of estimating speckle and far-field correlation functions of visible light in terms of the diameter ($D=2a$), this may not be a serious limitation because within the visible spectrum the sun and presumably many other stars exhibit to a first-order approximation the uniform intensities of Lambertian radiators.  In any case, as mentioned above and explained in Section \ref{subsec:Lambertian}, known limb darkening/brightening functions can be included in the analysis if required.
\par
In summary, we model the sun and other stars in a narrow received visible bandwidth by a spherical antenna of radius $a$ with a thin uniform surface layer of statistically independent closely spaced (average separation distance of about a half wavelength or less) volume sources that produce Lambertian radiation.
\par
By using the classical macroscopic Maxwell equations to represent the fields of a star, it is assumed that once the photons leave the molecules of the star and enter the free-space region $r\ge a$, their quantum nature is subsumed in the space-time-average classical wave behavior of the fields produced by the myriad of photons in free space.  Historically, there was some question as to whether the quantum effects of the photoelectric detectors used to measure the fields in intensity interferometry would significantly change the classically derived correlation functions.  However, Purcell showed that the statistically averaged counting rate of the photoelectrons indeed produces the classically predicted cross correlation in the intensity of the incident electric field \cite{Purcell}, \cite[secs. 7.3--7.4]{Wolf}.
\section{Terminology, Concepts, and Methodology Explained Using a Line-Source Radiator}\label{sec:TCM}
As a way of defining and explaining some of the important terminology, concepts, and methods used in this paper, we begin with  a simple example of a scalar-field (acoustic) radiator composed of single-frequency (monochromatic) delta-function point sources with coefficients $4\pi s_i(t)$ located at fixed points $\vrho_i$ separated by distances much less than a wavelength along a straight line from $-a$ to $+a$, such that
\be{P1}
s_i(t) = A_i\cos[\w_0 t + \psi_i]
\ee
with $\w_0$ equal to the single frequency of all the point sources, whereas both the magnitude $A_i$ and phase $\psi_i$ of each point source can be different within their ranges $0\le A_i \le A_{\rm max}$ and $0\le \psi_i < 2\pi$.  The $A_i$ and  $\psi_i$ of each point source can represent the average magnitude and phase of an electrically small incremental length of a continuous line-source radiator. The scalar field $E(\vr,t)$ produced by the point sources sa\-tis\-fies the Helmholtz equation and thus can be expressed as
\bea{P2}
E(\vr,t) =  \sum_i A_i\cos[\w_0( t -|\vr-\vrho_i|/c) + \psi_i]/|\vr-\vrho_i|\nonumber\\  = A(\vr)\cos[\w_0 t + \psi(\vr)]
\eea
where $A(\vr)$ is a positive magnitude and $c$ is the free-space speed of light.  \textit{Note that we are using the real-valued sources and fields and not the complex phasor or analytic-signal sources and fields.}
\par
The sources and fields in (\ref{P1}) and (\ref{P2}) can be said to be \textit{coherent} for all $\vr$ and $t$ (phase differences between all points are the same for all time) if and only if the $A_i$ and $\psi_i$ (and thus the $A(\vr)$ and $\psi(\vr)$) are independent of time \cite[sec. 4.2]{B&P}.  Monochromatic fields are coherent for any set of values of the magnitudes and phases of the sources provided the magnitudes and phases are constant (independent of time).  Of course, all radiators have a finite bandwidth.  However, for most ordinary antennas tuned to a particular frequency within the bandwidth, the sources and fields can be characterized by constant magnitudes and phases at that frequency, and thus the sources and fields of such antennas can be considered coherent.  If the $A_i$ or $\psi_i$ (and thus the $A(\vr)$ or $\psi(\vr)$) depend on time, then the fields are not coherent throughout all space and time but they may be \textit{partially coherent} in that they have slowly varying phase and magnitude compared to $\w_0 t$ and $|\cos(\w_0 t)|$, respectively, over portions of space and time, such as within the wave packets of quasi-monochromatic fields.
\subsection{Quasi-Monochromatic Fields of Line Sources}\lbl{sub1:Quasi}
For sources and fields with a finite but narrow observed bandwidth $\Delta\w_0$ such that $\Delta \w_0/\w_0 \ll 1$, the Fourier transform of the frequency-domain source coefficients and fields can be used to show (see Section \ref{subsec:Quasi} below) that ``quasi-monochromatic'' time-domain source coefficients and fields replace those in (\ref{P1}) and (\ref{P2}), namely
\be{P3}
s_i(t) = A_i(t)\cos[\w_0 t + \psi_i(t)]
\ee
\be{P4}
E(\vr,t) =  \sum_i A_i(t')\cos[\w_0 t' + \psi_i(t')]/|\vr-\vrho_i|  
\ee
where $t' = t -|\vr-\vrho_i|/c$.  Although the radiation from the sun and most stars is broadband, it is assumed that the measurement system filters the received fields to a narrow (quasi-monochromatic) bandwidth.   The amplitude (magnitude) and phase modulation functions, $A_i(t)$ and $\cos\psi_i(t)$, have a minimum time period equal to about $2\pi/\Delta\w_0$, which is much longer than the monochromatic time period of $2\pi/\w_0$.   (If the period of the modulation functions were appreciably shorter than $2\pi/\Delta\w_0$, then we would have the contradictory result that the bandwidth would be appreciably larger than $\Delta\w_0$.)  The $A_i(t)$ and $\cos\psi_i(t)$ are bandlimited modulation functions that continue indefinitely in time but never repeat.  That is, each of the quasi-monochromatic source coefficients has a time dependence consisting of a continual nonrepeating sequence of envelopes of a modulated sinusoidal wave with mean carrier frequency $\w_0$.  The minimum narrow-band pulse (envelope) width is about equal to $2\pi/\Delta\w_0$.  For star-like sources, time pulses having a minimum width approximately equal to $2\pi/\Delta\w_0$ come one right after the other because there are a myriad of statistically independent molecular sources within a star that are continually emitting radiation to fill the observed bandwidth $\Delta\w_0$.  Narrow bandwidth fields recorded for a long duration of time $T\gg 2\pi/\Delta\w_0$ have a frequency-spectrum magnitude and phase that can vary rapidly with frequency over this narrow bandwidth $\Delta\w_0$.  Moreover, the exact magnitude and phase variation depends on the length of the observation time $T$.
\par
\begin{figure}[!t]
\centerline{\includegraphics[width=6cm]{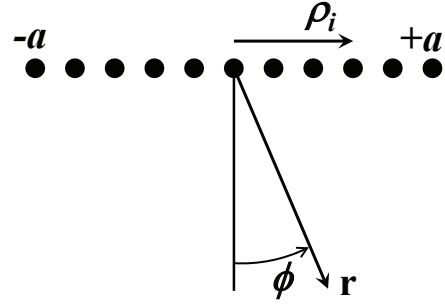}}
\vspace{-2mm}
\caption{Delta-function point sources separated by distances much less than a wavelength in a linear array along a line from $-a$ to $+a$.}
\label{figlinesource}
\vspace{-3mm}
\end{figure}

The behavior of the electric field in (\ref{P4}) can be determined in greater detail by expanding $|\vr -\vrho_i|$ in the following power series, which is convergent for $r\ga 2a$
\be{PP}
|\vr -\vrho_i| = r\left[1 -\frac{\rho_i\sin\phi}{r} + O(\frac{\rho^2_i\sin^2\phi}{r^2}) + O(\frac{\rho^2_i}{r^2})\right]
\ee
where $r=|\vr|$,  $\rho_i = \pm|\vrho_i|$ with the $\pm$ signs chosen if $\vrho_i$ is on the $\pm a$ side of the line, and $\phi$ is the angle between $\vr$ and the normal to the line of sources in a plane containing the line of sources, as shown in Fig. \ref{figlinesource}.  Then (\ref{P4}) can be rewritten as
\bea{P4a}
&&\hspace{-7mm}E(\vr,\!t) \!=\! \frac{1}{r} \!\!\sum_i\!\! A_i\!\left(\!t\!-\!\frac{r}{c}\!\!\left[1\! -\!\frac{\rho_i\sin\phi}{r}\! +\! O(\frac{\rho^2_i\sin^2\phi}{r^2})\!+\! O(\frac{\rho^2_i}{r^2})\right]\!\right)\nonumber\hspace{-2mm}\\ 
&&\hspace{-7mm}\cdot\cos\left\{\w_0 \left(t-\frac{r}{c}\left[1 -\frac{\rho_i\sin\phi}{r} + O(\frac{\rho^2_i\sin^2\phi}{r^2}) + O(\frac{\rho^2_i}{r^2})\right]\right)\right.\nonumber\\ &&\hspace{-8mm}\left.+ \psi_i\left(t-\frac{r}{c}\left[1 -\frac{\rho_i\sin\phi}{r} + O(\frac{\rho^2_i\sin^2\phi}{r^2}) + O(\frac{\rho^2_i}{r^2})\right]\right)\right\}.\hspace{-3mm} 
\eea
If we expand the functions $A_i$ and $\psi_i$ in a power series about $t-r/c$ and note that $dA_i(t)/dt = 0(\Delta\w_0)A_i(t)$ and $d\psi_i(t)/dt = 0(\Delta\w_0)\psi_i(t)$, then (\ref{P4a}) becomes 
\bea{P4b}
&&\hspace{-7mm}E(\vr,t) = \frac{1}{r} \sum_i \left[1+0(\Delta\w_0 \rho_i\sin\phi/c)+0(\Delta\w_0 \rho^2_i/(rc))\right]\nonumber\\[-3mm]&&\hspace{50mm}\cdot A_i(t-r/c)\nonumber\\ 
&&\hspace{-7mm}\cdot\cos\bigg\{\w_0 \left(t-\frac{r}{c}\left[1 -\frac{\rho_i\sin\phi}{r} + O(\frac{\rho^2_i\sin^2\phi}{r^2}) + O(\frac{\rho^2_i}{r^2})\right]\right)\nonumber\\ &&\hspace{-8mm}+\! \left[1+0(\Delta\w_0 \rho_i\sin\phi/c)+0(\Delta\w_0 \rho^2_i/(rc))\right]\psi_i(t-r/c)\bigg\}\!.  
\eea
For $r\ga 2a$, the predominant variation with $\phi$ and $r$ in the terms of the summation in (\ref{P4b}) is given by $\cos[k_0(r-\rho_i\sin\phi + O(\rho_i^2/r))]$.  The $\phi$ part of this variation, which is independent of $r$ and $t$, produces the far-field pattern of a $2a$ wide linear array of sources with random excitations.  Such an array has far-field lobes with an average beamwidth given by changes in $k_0a\sin\phi$ on the order of $\pi$.  Since $\Delta\w_0/\w_0 \ll 1$, the corresponding change in $\Delta\w_0\rho_i\sin\phi/c$ is negligible ($\ll 1$) and there is no change with $\phi$ in the $0(\Delta\w_0 \rho^2_i/(rc))$ terms.  Thus, the $[0(\Delta\w_0 \rho_i\sin\phi/c)+0(\Delta\w_0 \rho^2_i/(rc))]$ contributions as well as the $O(\rho^2_i\sin^2\phi/r^2)$ contributions in (\ref{P4b}) can be omitted without changing the statistical character of the local $\phi$ variations of the field.
\par
The average radial length of the envelopes of the wave packets comprising the field are determined by the $A_i(t-r/c)\cos[\psi_i(t-r/c)]$ functions.  Since the bandwidth of the pulses is $\Delta\w_0$, and the group speed of a wave packet in free space equals $c$, this average radial length is on the order of the minimum radial length  $\Delta r = 2\pi c/\Delta\w_0$ of the wave packets.  As $r$ changes by $\Delta r$, the  $O(\rho^2_i/r^2)$ terms change the argument of the cosine functions in (\ref{P4b}) by an amount given approximately as $k_0 a^2[1/r-1/(r+\Delta r)]$.  Therefore, the $O(\rho^2_i/r^2)$ in the cosine functions of (\ref{P4b}) can be neglected without changing the statistical character of the local variation with $r$ provided
\be{crit}
r\ga 2\,{\rm Min}\!\left[a\sqrt{\w_0/\Delta\w_0}\,,a^2/\lambda_0\right].
\ee
Likewise, under the same criterion in (\ref{crit}), the $O(\Delta\w_0\rho^2_i/(rc))$ contributions in (\ref{P4b}) can be neglected without changing the essence of the local $r$ variations of the field.
\par
Consequently, without changing the predominant local $\phi$ and $r$ variations in the field, all the $O(\cdots)$ terms in (\ref{P4b}) can be omitted under the condition (\ref{crit}), so that (\ref{P4b}) reduces to
\be{P4c}
E(\vr,t)\! = \!\frac{1}{r} \sum_i\! A_i(t-\frac{r}{c})\cos[\w_0 (t-\frac{r}{c}) + k_0\rho_i\sin\phi + \psi_i(t-\frac{r}{c})].  
\ee
Rewriting (\ref{P4c}) as 
\be{P4d}
E(\vr,t)\! =\! \frac{1}{r}{\rm Re}\Big[ e^{i\w_0(t-\frac{r}{c})}\!\sum_i\! A_i(t-\frac{r}{c}) e^{i[\psi_i(t-\frac{r}{c}) +k_0a(\frac{\rho_i}{a})\sin\phi]}\Big]  
\ee
shows that the electric field $E(\vr,t)$ can be expressed as
\be{P4e}
\frac{1}{r} A[(t-\frac{r}{c}),k_0a\sin\phi]\! \cos\{\w_0(t-\frac{r}{c})\! +\! \psi[(t-\frac{r}{c}),k_0a\sin\phi]\}
\ee
where the $A$ and $\psi$ are positive-magnitude (amplitude) and phase functions of both the variables $(t-r/c)$ and $k_0a\sin\phi$.
\begin{figure}[!t]
\centerline{\includegraphics[width=\columnwidth]{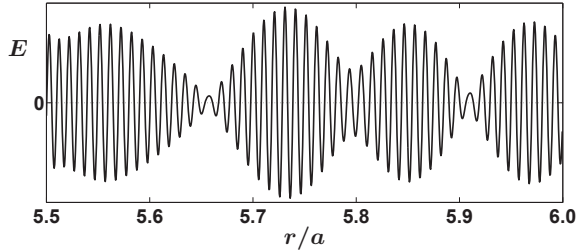}}
\vspace{-5mm}
\caption{Representative snapshot in time of the wave train of the electric field as a function of the radius $r$  at a fixed angle $\phi$ for the linear array of incoherent sources in Fig. \ref{figlinesource} with $a = 100$ meters, $\lambda_0 = 1$ meter, and a bandwidth $\Delta\w_0/\w_0 = 0.1$.  The incoherent sources are equally spaced with a separation distance of $\lambda_0/5$ and were given random magnitudes between $0$ and $1$ and random phases between $0$ and $2\pi$.}
\label{figwavetrain}
\vspace{-5mm}
\end{figure}
\begin{figure}[!t]
\centerline{\includegraphics[width=\columnwidth]{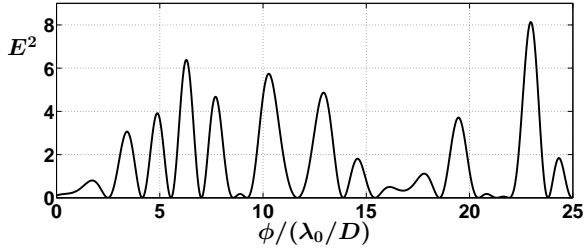}}
\vspace{-6mm}
\caption{Representative snapshot in time of the electric-field intensity pattern as a function of angle $\phi$ at the fixed radius $r = 5a$ for the same linear array of incoherent sources described in the caption of Fig. \ref{figwavetrain}.} 
\label{figfieldpattern}
\vspace{-5mm}
\end{figure}
\par
At any fixed field angle $\phi$, (\ref{P4e}) is simply a train of wave packets with carrier frequency $\w_0$ and a minimum \textit{wave-packet length} of about $2\pi c/\Delta \w_0$, provided $\Delta \w_0/\w_0 \ll 1$.  Both the phase velocity of the carrier wave and group velocity of the wave packets are in the same direction and have the same magnitude equal to $c$.  A representative snapshot in time of the radial train of wave packets at a fixed angle $\phi$ was computed for the linear array of incoherent sources shown in Fig. \ref{figlinesource} and is plotted in Fig. \ref{figwavetrain}.  The numerical results depicted in Fig. \ref{figwavetrain} confirm the theoretical predictions that the carrier wavelength is approximately equal to the center wavelength $\lambda_0$ and that the minimum wave-packet length is approximately $2\pi c/\Delta\w_0 = (\w_0/\Delta\w_0)\lambda_0$.  In addition, the numerical simulations confirmed the criterion in (\ref{crit}), namely that the radial train of wave packets become more chaotic and less predictable for $r\la 2a\sqrt{\w_0/\Delta\w_0}$\,.
\par
Note that the far-field behavior expressed in (\ref{P4c})--(\ref{P4e}) of the quasi-monochromatic wave trains emitted by the incoherent sources is valid under the criterion in (\ref{crit}), which, for typical stellar interferometer measurements that have $a/\lambda_0 \gg \sqrt{\w_0/\Delta\w_0}$, is much less restrictive than the far-field criterion of $r\ga 2a^2/\lambda_0$ (Rayleigh distance) that applies to monochromatic waves emitted by coherent sources ($\Delta\w_0/\w_0\to 0$).  Moreover, with regard to changes in $\phi$ for a fixed $r$, it was shown above that the criterion in (\ref{crit}) for (\ref{P4c})--(\ref{P4e}) to hold can be replaced by $r\ga 2a$.  For a sphere (rather than a line) of incoherent sources, power conservation and spherical symmetry would indicate that a similar far-field $\phi$ variation would hold everywhere outside the reactive zone of the sphere, that is, for $r$ greater than a few wavelengths from the surface of the sphere.     Agarwal, Gbur, and Wolf \cite{AGW} showed that this far-field $\phi$ variation produced by incoherent sources outside the reactive zone holds also for the numerically computed cross-spectral density function of the fields radiated by a spherical star under the assumption that the fields at the surface of the star exhibit delta-function correlation.  
\par
One way to get an idea of how rapidly the electric field varies with the angle $\phi$ for a fixed radius $r$ and fixed time $t$ is to note in (\ref{P4c}) that $\cos(k_0|\rho_i|\sin\phi)$ encounters consecutive zeros and peaks as $k_0|\rho_i|\sin\phi$ changes by $\pi/2$.  If we insert an average $|\rho_i|=a/2$, this implies that the lobes of the field pattern versus $\phi$ would have an average half-power beamwidth of about $\lambda_0/(2a)=\lambda_0/D$.  A representative snapshot in time of the far-field intensity pattern versus angle $\phi$ at a fixed radius $r$ was also computed for the linear array of incoherent sources shown in Fig. \ref{figlinesource} and is plotted in Fig. \ref{figfieldpattern}.  The numerical results depicted in Fig. \ref{figfieldpattern} confirm the theoretical predictions that the average half-power beamwidth of the lobes of the far-field pattern is equal to about $\lambda_0/D$.   Moreover, the numerical simulations confirmed that this representative angular electric-field pattern as a function of $\phi$ with average half-power beamwidth equal to about $\lambda_0/D$ holds for all $r\ga 2a$.
\par  
This far-field intensity pattern versus $\phi$ shown in Fig. \ref{figfieldpattern} is called the \textit{angular speckle pattern} of the incoherent line sources.   For three-dimensional volume sources, the angular speckle pattern would be defined for two far-field angles $(\theta,\phi)$.  (``Speckle interferometry'' \cite{Labeyrie} is a method for determining the diameter of stars by laser processing the speckle pattern observed in short exposures with a single large telescope in order to directly measure the average speckle size.)   As we shall show next, a useful quantitative measure of the speckle pattern can be obtained by defining and evaluating correlation functions for the fields.
\subsection{First-Order Temporally Averaged Correlation Functions for Line Sources}\lbl{sub1:FOT}
The temporally averaged, first-order correlation function is defined for real-valued quasi-monochromatic source coefficients as\\[-5mm]
\be{P5}
G^{(1)}_{ij}(\tau) = \langle s_i(t+\tau) s_j(t) \rangle\\[-2mm]
\ee
for all source points $(i,j)$ and time differences $\tau$, where $\langle\;\rangle$ denotes the time average over a time $T$ long enough that the correlation function changes negligibly for longer times, namely $T\gg 2\pi/\Delta\w_0$.  In principle, the time $T$ can be allowed to approach infinity so that the time average of a function $f(t)$ can be rigorously defined as\\[-3mm]
\be{P6}
\langle f(t) \rangle = \lim_{T\to\infty}\frac{1}{T}\int\limits_0^{T} f(t) dt\\[-2mm]
\ee
where the averaging begins at time $t=0$.
\par
Similarly, the temporally averaged, first-order correlation function for the real quasi-monochromatic fields for two ob\-ser\-va\-tion points $(\vr_p,\vr_q)$ and time difference $\tau$ is defined as\\[-4mm]
\be{P7}
G^{(1)}(\vr_p,\vr_q,\tau) = \langle E(\vr_p,t+\tau) E(\vr_q,t) \rangle.\\[-2mm]
\ee
For $i\neq j$ and $p\neq q$, the $G^{(1)}_{ij}(\tau)$ and $G^{(1)}(\vr_p,\vr_q,\tau))$ are commonly referred to as \textit{cross-correlation functions}. 
For $i=j$ and $p=q$, they are commonly called \textit{autocorrelation functions} \cite[sec. 2.4]{Wolf}.
\par
Insertion of the quasi-monochromatic source coefficients from (\ref{P3}) into (\ref{P5}) gives\\[-5mm]
\bea{P8}
G^{(1)}_{ij}(\tau) = \langle A_i(t+\tau) A_j(t)\hspace{35mm}\nonumber\\\cdot\cos[\w_0 (t+\tau) + \psi_i(t+\tau)]\cos[\w_0 t + \psi_j(t)] \rangle.
\eea
\mbox{}\\[-6mm]
Rewriting the product of cosines in (\ref{P8}) as $\{\cos[2\w_0 t +\w_0\tau + \psi_i(t+\tau)+\psi_j(t)] + \cos[\w_0\tau+\psi_i(t+\tau)-\psi_j(t)]\}/2$ shows immediately that the first cosine term in this expression time averages to zero, reducing (\ref{P8}) to\\[-3mm]
\be{P10}
 G^{(1)}_{ij}(\tau) =\textstyle\frac{1}{2}\langle A_i(t+\tau) A_j(t)\cos[\w_0\tau + \psi_i(t+\tau)- \psi_j(t)] \rangle.
\ee
For the statistically independent molecular sources of a star, the bandlimited phases $\psi_i(t)$ and $\psi_j(t)$ of the effective source coefficients are independent functions of time (for $i\neq j$) that continuously vary over the full range of phase from $0$ to $2\pi$ with each of the values of phase in this full range occurring with equal probability.  Thus, the long-term time average of the cosine function in (\ref{P10}) multiplied by the positive magnitudes $A_i(t+\tau) A_j(t)$ will also approach zero for averages measured over times $T \gg 2\pi/\Delta\w_0$ in (\ref{P6}).  In other words, $G^{(1)}_{ij}(\tau)$ equals zero for $i\neq j$, so that
\be{P11}
G^{(1)}_{ij}(\tau) = A^2_{0i}(\tau)\delta_{ij}\\[-4mm]
\ee
where
\be{P11'}
A^2_{0i}(\tau)=\langle A_i(t+\tau) A_i(t)\cos[\psi_i(t+\tau)- \psi_i(t)]\rangle/2
\ee
with $\w_0\tau$ omitted because it has no effect on the time average.
The cross-correlation in (\ref{P11}) for all the different quasi-monochromatic source coefficients of the star are zero and the sources can be said to be \textit{incoherent}; the source coefficients have Kronecker-delta correlation.  ``Incoherence" is defined as having zero cross correlation.  It can be proven in general that statistically independent random variables have zero cross correlation (but zero cross correlation does not necessarily imply statistical independence).
\par
Similarly, insertion of the quasi-monochromatic electric field from (\ref{P4c}) into (\ref{P7}) gives
\bea{P12}
G^{(1)}(\vr_p,\vr_q,\tau) =  \frac{1}{r^2}\sum_{i,j}\langle A_i(t +\tau)A_j(t)\cos[\w_0 (t+\tau)\\ + k_0\rho_i\sin\phi_p + \psi_i(t+\tau)]\cos[\w_0 (t) + k_0\rho_j\sin\phi_q + \psi_j(t)] \rangle\nonumber\hspace{-7mm}
\eea
with $r_p = r_q = r$ and the $r/c$ omitted in $(t-r/c)$ because it has no effect on the time average.
Re-expressing the product of cosines in (\ref{P12}), and noting that the time average of the  cosine term with $2\w_0 t$ in its argument is zero, we find that (\ref{P12}) reduces to
\bea{P13}
&&\hspace{-7mm}G^{(1)}(\vr_p,\vr_q,\tau) = \frac{1}{2r^2} \sum_{i,j}\langle A_i(t +\tau)A_j(t)\\&&\hspace{-6mm}\cdot\cos[\w_0\tau +k_0(\rho_i\sin\phi_p - \rho_j\sin\phi_q) + \psi_i(t+\tau)- \psi_j(t)] \rangle.\nonumber
\eea
Again, due to the independent, uniformly distributed, random time variation of the $\psi_i(t)$ and $\psi_j(t)$ phase functions, all the terms in (\ref{P13}) with $i\neq j$ have long-term time averages that approach zero for $T\gg 2\pi/\Delta\w_0$.  Thus, (\ref{P13}) reduces to
\bea{P14}
&&\hspace{-7mm}G^{(1)}(\vr_p,\vr_q,\tau) = \frac{1}{2r^2} \sum_i\langle A_i(t +\tau)A_i(t)\\&&\hspace{-6mm}\cdot\cos[\w_0\tau +k_0\rho_i(\sin\phi_p - \sin\phi_q) + \psi_i(t+\tau)- \psi_i(t)] \rangle\nonumber
\eea
which further simplifies to
\be{P14a}
G^{(1)}(r,\Delta\phi,0) = \frac{1}{2r^2} \sum_i\langle A^2_i(t)\rangle\cos[k_0\rho_i\sin\Delta\phi] 
\ee
with $\tau=0$, the initial angle $\phi_q=0$, and $\phi_p= \Delta\phi$.
\par
For many closely spaced sources, the summation in (\ref{P14a}) is well approximated by the integral\\[-3mm] 
\be{P14b}
G^{(1)}(r,\Delta\phi,0) = \frac{1}{2r^2} \int\limits_{-a}^{+a}\langle {\cal A}^2(\rho,t)\rangle\cos[k_0\rho\sin\Delta\phi]\, d\rho\\[-2mm]
\ee
where ${\cal A}^2(\rho,t)$ is the continuous representation of $A_i^2(t)$ normalized to the average increments between the source points.  If ${\cal A}^2(\rho,t)$ is independent of the position from $-a$ to $+a$ (no intensity taper) such that $\langle{\cal A}^2(\rho,t)\rangle ={\cal A}^2_0$, then\\[-2mm]
\be{P15}
G^{(\!1\!)}(r,\!\Delta\phi,\!0)\! = \! \frac{{\cal A}_0^2}{r^2}\!\!\int\limits_{0}^{+a}\!\!\cos[k_0\rho\sin\Delta\phi]\, d\rho\! =\! \frac{\sin({\cal A}_0^2k_0a\sin\Delta\phi)}{r^2k_0\sin\Delta\phi}\!.
\ee
Note from (\ref{P14a}) and (\ref{P15}) that\\[-3mm]
\be{P15'}
\sum_i\langle A^2_i(t)\rangle = 2a{\cal A}_0^2.\\[-2mm]
\ee
The corresponding normalized, temporally averaged, first-order correlation function for the real-valued quasi-monochromatic fields is thus
\be{P16}
g^{(1)}(r,\Delta\phi,0) = \frac{G^{(1)}(r,\Delta\phi,0)}{G^{(1)}(r,0,0)} = \frac{\sin(k_0a\sin\Delta\phi)}{k_0a\sin\Delta\phi}.
\ee
The first zero of this first-order correlation function occurs at 
\be{P17}
\Delta\phi_0 = \sin^{-1}\frac{\lambda_0}{2a} \approx \frac{\lambda_0}{2a} = \frac{\lambda_0}{D}.
\ee
This indicates that the average half-power beamwidth of the lobes of the speckle pattern in the $\phi$ direction for these incoherent line sources is about $\lambda_0/D$, which is equal to the half-power beamwidth of the main lobe of the far-field pattern of a uniform phase and magnitude/amplitude coherent line-source radiator.  Thus, we see that the cross correlation of the fields at two different points in space, unlike the cross correlation of the different source coefficients, is not equal to zero.  The fields are \textit{partially coherent} but not incoherent even though the sources are incoherent.
\subsection{Second-Order Temporally Averaged Correlation Function for Line Sources}\lbl{sub:SOC}
High tolerances required by Michelson stellar interferometry with first-order temporally averaged correlation functions encouraged the development by Hanbury Brown--Twiss in the 1950's of ``intensity interferometry'' \cite{HBT1956}, \cite{HBT1957} that required only the correlation of the magnitudes (intensities) of the electric field measured by two different photoelectric detectors to obtain the diameters of stars.
\par
For intensity interferometry, the response time $T_d$ of the photoelectric detectors are much longer than the period $2\pi/\w_0$ of the mean (carrier) frequency and much shorter than the period $2\pi/\Delta\w_0$ of the modulation frequency, that is
\be{P67}
2\pi/\w_0 \ll T_d \ll 2\pi/\Delta\w_0
\ee
which implicitly requires $\Delta\w_0/\w_0 \ll 1$.  Placing the first photoelectric detector at the angle $\phi_1$, it measures the running average $\overline{E^2(r,\phi_1,t)}$ (denoted by the overline) over the response time $T_d$ of the detector.   We have from (\ref{P4c}) that
\bea{P68}
&&\hspace{-6mm}\overline{E^2(r,\phi_1,t)} \!=\! \frac{1}{r^2}\!\!\sum_{i,j}\!\! A_i(t)A_j(t)\overline{\cos[\w_0 (t) \!+\! k_0\rho_i\sin\phi_1 \!+\! \psi_i(t)]}\nonumber\\&&\hspace{25mm}\cdot\overline{\cos[\w_0 (t) + k_0\rho_j\sin\phi_1 + \psi_j(t)}\,. 
\eea
The time variation of $A_i(t)$ has an average period of about $2\pi/\Delta\w_0\! \gg \!T_d$ so $\overline{A_i(t)A_j(t)}$ has been set equal to $A_i(t)A_j(t)$.  If the product of the cosines in (\ref{P68}) is rewritten as the sum of two cosines with their arguments equal to the sum and dif\-fer\-ence of the arguments of the cosines in the product, then the cosine term with argument $2\w_0 t$ averages to zero, leaving\\[-5mm] 
\bea{P69}
&&\hspace{-10mm}\overline{E^2(r,\phi_1,t)} = \frac{1}{2r^2}\sum_{i,j} A_i(t)A_j(t)\nonumber\\[-1mm]&&\cdot\cos[k_0(\rho_i-\rho_j)\sin\phi_1 + \psi_i(t)-\psi_j(t)]. 
\eea
\mbox{}\\[-5mm]
Like $A_i(t)$, the time variation of the cosine term in (\ref{P69}) has an average period of about $2\pi/\Delta\w_0 \gg T_d$ so the overline denoting the running time average can be omitted.  Similarly, for the angle $\phi_2$ we have\\[-4mm]
\bea{P70}
&&\hspace{-10mm}\overline{E^2(r,\phi_2,t)} = \frac{1}{2r^2}\sum_{i,j} A_i(t)A_j(t)\nonumber\\&&\cdot\cos[k_0(\rho_i-\rho_j)\sin\phi_2 + \psi_i(t)-\psi_j(t)].
\eea
\mbox{}\\[-9mm]
\par
If the output in (\ref{P69}) of the photoelectric detector at $\phi_1=0$ is correlated with the output in (\ref{P70}) of the photoelectric detector at $\phi_2 = \Delta\phi$ for a long-time average $T\gg 2\pi/\Delta\w_0 \gg T_d$ (see (\ref{P6})), we obtain the second-order temporally averaged correlation function
\be{P71}
G^{(2)}(r,\Delta\phi,0) = \left\langle\overline{E^2(r,0,t)}\;\overline{E^2(r,\Delta\phi,t)}\right\rangle.
\ee
Substitution from (\ref{P69}) and (\ref{P70}) into (\ref{P71}) yields
\bea{P72}
&&\hspace{-7mm}G^{(2)}(r,\Delta\phi,0) = \frac{1}{4r^4}\sum_{i,j}\sum_{i',j'} \left\langle A_i(t)A_j(t)A_{i'}(t)A_{j'}(t)\right.\\&&\left.\hspace{-8mm}\cdot\cos[\psi_i(t)\!-\!\psi_j(t)]\cos[k_0(\rho_{i'}\!-\!\rho_{j'})\sin\Delta\phi \!+ \!\psi_{i'}(t)\!-\!\psi_{j'}(t)]\right\rangle.\nonumber 
\eea
For $i = j$, the long-time average approaches zero unless $i' = j'$ because the phase $[\psi_{i'}(t)-\psi_{j'}(t)]$ continuously varies with time between $0$ and $2\pi$ except for $i' = j'$.  The contribution to the quadruple summation in (\ref{P72}) from $[i = j, i' = j']$ for point sources with no intensity taper is found from (\ref{P15'}) as\\[-3mm]
\be{P73}
\Big(\sum_i \left\langle A^2_i(t)\right\rangle\Big)^2  = (2a {\cal A}^2_0)^2\\[-2mm]
\ee
which allows (\ref{P72}) to be re-expressed as\\[-3mm]
\bea{P74}
&&\hspace{-6mm}G^{(2)}(r,\Delta\phi,0) = \frac{(a {\cal A}^2_0)^2}{r^4}\nonumber\\ &&\hspace{-7mm}+ \frac{1}{4r^4}\sum\limits_{i,j}^{i\neq j}\sum\limits_{i',j'}^{i'\neq j'}\left\langle A_i(t)A_j(t)A_{i'}(t)A_{j'}(t)\cos[\psi_i(t)-\psi_j(t)]\right.\nonumber\\ &&\hspace{-7mm}\left.\cdot\cos[k_0(\rho_{i'}-\rho_{j'})\sin\Delta\phi + \psi_{i'}(t)-\psi_{j'}(t)]\right\rangle.
\eea
\mbox{}\\[-5mm]
The product of the cosines in (\ref{P74}) can be rewritten as
\bea{P75}
&&\hspace{-9mm}\textstyle\frac{1}{2}\!\cos[k_0(\rho_{i'}\!-\!\rho_{j'}\!)\!\sin\Delta\phi\!+\! \psi_i(t)\!-\!\psi_j(t)\!+\! \psi_{i'}(t)\!-\!\psi_{j'}(t)]\hspace{10mm}\hspace{-10mm}\\
&&\hspace{-8mm}+\textstyle\frac{1}{2}\!\cos[k_0(\rho_{i'}\!-\!\rho_{j'})\!\sin\Delta\phi\!-\! \psi_i(t)\!+\!\psi_j(t)\!+\! \psi_{i'}(t)\!-\!\psi_{j'}\!(t)].\nonumber
\eea
The long-time average in (\ref{P74}) of the second cosine function in (\ref{P75}) approaches zero unless $i=j'$ and $i'=j$ such that this cosine function equals\\[-3mm]
\be{P76}
\cos[k_0(\rho_{i'}-\rho_{j'})\sin\Delta\phi]/2.\\[-1mm]
\ee
By a similar argument, the last cosine function in (\ref{P75}) also reduces to (\ref{P76}).  In all then, the quadruple summation in (\ref{P74}) reduces to a double summation and $G^{(2)}(r,\Delta\phi,0)$ becomes\\[-3mm]
\be{P77}
\frac{(a {\cal A}^2_0)^2}{r^4}\! + \!\frac{1}{4r^4}\sum\limits_{i,j}^{i\neq j}\!\left\langle A^2_i(t)A^2_j(t)\right\rangle\cos[k_0(\rho_i\!-\!\rho_j)\sin\Delta\phi].\\[-2mm]
\ee
For point sources with uniform spacing $\Delta\rho$, the $A^2_i(t)$ are all equal to the same constant plus a function with time average equal to zero. Thus, we have from (\ref{P73}) that
\be{P78}
\left\langle A^2_i(t)A^2_j(t)\right\rangle  =  (\Delta\rho {\cal A}^2_0)^2,\;\;\;i\neq j
\ee
and (\ref{P77}) reduces to\\[-3mm]
\be{P79}
G^{(2)}(r,\Delta\phi,0) = \frac{a^2 {\cal A}_0^4}{r^4} + \frac{\Delta\rho^2 {\cal A}_0^4}{4r^4}\sum\limits_{i,j}^{i\neq j}\!\cos[k_0(\rho_i-\rho_j)\sin\Delta\phi].
\ee
\par
Rewriting the cosine function in (\ref{P79}) as
\bea{P79'}
\cos(k_0\rho_i\sin\Delta\phi)\cos(k_0\rho_j\sin\Delta\phi)\hspace{20mm}\nonumber\\ + \sin(k_0\rho_i\sin\Delta\phi)\sin(k_0\rho_j\sin\Delta\phi)
\eea
(\ref{P79}) can be expressed as
\bea{P80}
&&\hspace{-7mm}G^{(2)}(r,\Delta\phi,0) = \frac{a^2 {\cal A}_0^4}{r^4} + \frac{\Delta\rho^2 {\cal A}_0^4}{4r^4}\bigg(\Big(\sum\limits_i \cos(k_0\rho_i\sin\Delta\phi)\Big)^2\nonumber\\[-1mm] &&\hspace{30mm}+ \Big(\sum\limits_i \sin(k_0\rho_i\sin\Delta\phi)\Big)^2\bigg).
\eea
Approximating the summations by integrals gives
\bea{P81}
\sum\limits_i \cos(k_0\rho_i\sin\Delta\phi) = \frac{1}{\Delta\rho}\int\limits_{-a}^{+a}\cos(k_0\rho\sin\Delta\phi)d\rho \nonumber\\[-1mm]= \frac{2a}{\Delta\rho}\frac{\sin(k_0a\sin\Delta\phi)}{k_0a\sin\Delta\phi}
\eea
\mbox{}\\[-7mm]
\be{P82}
\sum\limits_i \sin(k_0\rho_i\sin\Delta\phi) = \frac{1}{\Delta\rho}\int\limits_{-a}^{+a}\sin(k_0\rho\sin\Delta\phi)d\rho = 0
\ee
so that (\ref{P80}) reduces to
\bea{P83}
G^{(2)}(r,\Delta\phi,0) = \frac{a^2 {\cal A}_0^4}{r^4} \left[1+ \left(\frac{\sin(k_0a\sin\Delta\phi)}{k_0a\sin\Delta\phi}\right)^2\right]\nonumber\\ = \frac{a^2 {\cal A}_0^4}{r^4} \left(1+ \left[g^{(1)}(r,\Delta\phi,0)\right]^2\right).
\eea
Defining the normalized second-order temporally averaged correlation function as\\[-3mm]
\be{P84}
g^{(2)}(r,\Delta\phi,0) = \frac{G^{(2)}(r,\Delta\phi,0)}{\left\langle\overline{E^2(r,0,t)}\right\rangle^2}\\[-1mm]
\ee
and noting from (\ref{P7}) and (\ref{P15}) that
\be{P85}
\left\langle\overline{E^2(r,0,t)}\right\rangle^2 = \left[G^{(1)}(r,0,0)\right]^2 = \frac{a^2 {\cal A}_0^4}{r^4}
\ee
one finds the following simple relationship between the normalized first- and second-order correlation functions
\be{P86}
g^{(2)}(r,\Delta\phi,0) \!=\!1 + \left[\frac{\sin(k_0a\sin\Delta\phi)}{k_0a\sin\Delta\phi}\right]^2\!\!\! =\!  1 + \left[g^{(1)}(r,\Delta\phi,0)\right]^2\!\!\!.
\ee
This relationship between the first- and second-order correlation functions, which has been derived here directly and relatively effortlessly from the correlations of the time-domain fields, was utilized by Hanbury Brown and Twiss in their intensity stellar interferometric measurements.
\section{Spherical Mode Expansions for the Fields Outside a Spherical Star}\label{sec:SWE}
\begin{figure}[!t]
\centerline{\includegraphics[width=5cm]{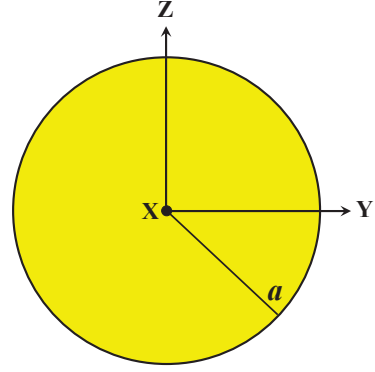}}
\caption{Model of spherical star with radius $a$ ($D=2a$), which lies in free space just outside the reactive fields of the sources of the star's radiation.}
\label{Figstar}
\vspace{-3mm}
\end{figure}

Consider the model discussed in Section \ref{subsec:Idealized} and shown in Fig. \ref{Figstar} of a spherical star with radius $a$, which lies in free space just outside the reactive fields of the sources of the star's radiation.
Rectangular coordinates $(x,y,z)$ with origin at the center of the sphere are also shown in Fig. \ref{Figstar}.  The  spherical coordinates $(r,\theta,\phi)$ of an observation point outside the sphere are defined in the usual way with respect to the $(x,y,z)$ axes; that is, the position vector $\vr$ makes an angle $\theta$ with the positive $z$ axis, and the projection of $\vr$ onto the $xy$ plane makes a right-hand angle $\phi$ with the positive $x$ axis.  The spherical angles $\theta$ and $\phi$ have the domains $[0,\pi]$ and $[0,2\pi]$, respectively.
\par
Let $\vE(\vr,t)$ denote the electric field of the star observed for a large but finite amount of time, for example, during a time interval $[0,T]$.  Taking the Fourier transform of $\vE(\vr,t)$ gives the frequency-domain electric field\footnote{Note that the electric-field functions $\vE(r,t)$ and $\vE_\w(\vr)$ depend on $T$.  However, if $T$ is much larger than the modulation period ($2\pi/\Delta\w_0$ -- see Sections \ref{sub1:Quasi} and \ref{subsec:Quasi}) of the wave packets emitted by the star, the essential statistical properties of $\vE(r,t)$ and $\vE_\w(\vr)$ are independent of $T$.}
\be{01}
\vE_\w(\vr) =\frac{1}{2\pi}\int\limits_0^T \vE(\vr,t)e^{i\w t} dt.
\ee
From Maxwell's equations, $\vE_\w(\vr)$ satisfies the homogeneous vector Helmholtz equation outside the sphere ($r\ge a$), namely
\be{1}
\nabla^2 \vE_\w (\vr) + k^2 \vE_\w(\vr) =0
\ee
where $k = 2\pi/\lambda = \w\sqrt{\mu_0\eps_0}=\w/c$ with $\lambda$ the wavelength and $\mu_0$ and $\eps_0$ equal to the permeability and permittivity of free space.  On the Earth where the fields of the star are measured, we have $r\gg a$ and the radial components of the fields are negligible compared to the $\theta$ and $\phi$ components.\footnote{For example, the radial component of the electric field can be proven negligible by writing $\nabla\cdot\vE_\w=0$ near the $xy$ plane ($\theta\approx\pi/2$) in spherical coordinates and noting that the radial variation in $\vE_\w(r,\theta,\phi)$ is on the order of $\exp(ikr)$ and the angular variation, as we show below, is no faster than $\exp(ika\theta)\exp(ika\phi)$.}  Thus, our main concern is to determine the $E_{\w\theta}$ and $E_{\w\phi}$ components from (\ref{1}).  With the help of \cite[ eq. (A2.99)]{vanBladel}, we can write the theta and phi components of (\ref{1}) as
\begin{subequations}
\lbl{2}
\bea{2a}
&&\hspace{-10mm}\nabla^2E_{\w\theta} +\frac{2}{r^2}\frac{\partial E_{\w r}}{\partial \theta} - \frac{E_{\w\theta}}{r^2 \sin^2\theta}
-\frac{2\cos\theta}{r^2\sin^2\theta}\frac{\partial E_{\w\phi}}{\partial \phi} + k^2 E_{\w\theta}\nonumber\\[-1mm]&&\hspace{60mm} = 0
\eea
\mbox{}\\[-11mm]
\bea{2b}
&&\hspace{-10mm}\nabla^2E_{\w\phi} +\frac{2}{r^2\sin\theta}\frac{\partial E_{\w r}}{\partial \phi} - \frac{E_{\w\phi}}{r^2 \sin^2\theta}
+\frac{2\cos\theta}{r^2\sin^2\theta}\frac{\partial E_{\w\theta}}{\partial \phi}\nonumber\\&&\hspace{40mm} + k^2 E_{\w\phi} = 0.
\eea
\end{subequations}
These equations can be shortened by noting that, as we show below, the spatial variations in the fields can be no faster than $\exp(ika\theta)\exp(ika\phi)$ and thus the $1/r^2$ terms in (\ref{2a}) and (\ref{2b}) are negligible compared with the $k^2 E_{\w\theta}$ and $k^2 E_{\w\phi}$ terms, respectively, since $ka\gg 1$ and $r\ge a$.  In other words,  to an extremely accurate approximation, $E_{\w\theta}$ and $E_{\w\phi}$ satisfy the homogeneous scalar Helmholtz equations 
\begin{subequations}
\lbl{4}
\be{4a}
\nabla^2E_{\w\theta} + k^2 E_{\w\theta} = 0
\ee
\be{4b}
\nabla^2E_{\w\phi} + k^2 E_{\w\phi} = 0.
\ee
\end{subequations}
Consequently, to determine either $E_{\w\theta}$ or $E_{\w\phi}$, we are left with solving the free-space frequency-domain equation
\be{5}
\nabla^2E_{\w}(r,\theta,\phi) + k^2 E_{\w}(r,\theta,\phi) = 0
\ee
for $r\ge a$, where $E_{\w}$ stands for either the $\theta$ or $\phi$ polarization of the frequency-domain electric field.
\par
\par
The solution to (\ref{5}) can be found in spherical coordinates in terms of a series of the spherical mode functions $h_n(kr) P_n^m(\cos\theta) e^{im\phi}$, where the $h_n(kr)$ are the first-kind spherical Hankel functions (for $e^{-i\w t}$ time dependence) and the $P_n^m(\cos\theta)$ are the associated Legendre polynomials \cite[sec. 9.6]{Jackson}.  However, it turns out that this representation is not well suited for determining correlation functions because the Legendre polynomials contain both the $n$ and $m$ indices.  A more suitable expansion is a complete Fourier series in both $\theta$ and $\phi$, that is
\be{6}
E_{\w}(r,\theta,\phi) = \sum\limits_{m=-\infty}^{+\infty}\sum\limits_{n=-\infty}^{+\infty} B_{nm}(\w,r)e^{in\theta}e^{im\phi}.
\ee
The right side of (\ref{6}) must satisfy the homogeneous Helm\-holtz equation in (\ref{5}).  Inserting (\ref{6}) into (\ref{5}) and using the orthogonality of the $e^{im\phi}$ on the $[0,2\pi]$ domain, we find
\bea{7}
\sum\limits_{n=-\infty}^{+\infty}\left[\frac{1}{r^2}\frac{\partial}{\partial r}\left(r^2\frac{\partial B_{nm}}{\partial r}\right)\hspace{40mm}\right.\nonumber\\  \left.+ \left(\frac{in}{r^2 \tan\theta} - \frac{1}{r^2}n^2 - \frac{m^2}{r^2\sin^2\theta} + k^2\right) B_{nm}\right]e^{in\theta} = 0.
\eea
This infinite-summation equation does not reveal a simple solution for $B_{nm}(\w,r)$ for all values of $\theta$ in its domain $[0,\pi]$ because of the presence of the $\tan\theta$ and $\sin^2\theta$ functions.   Fortunately, however, it can be solved for values of $\theta$ near $\pi/2$ where $\sin\theta\approx 1$ and $\cos\theta\approx 0$.  In this region of $\theta$, (\ref{7}) holds if $B_{nm}(\w,r)$ obeys the radial wave equation
\be{8}
\frac{1}{r^2}\frac{\partial}{\partial r}\left(r^2\frac{\partial B_{nm}}{\partial r}\right) - \frac{1}{r^2}(n^2 + m^2)B_{nm} + k^2 B_{nm} =0.
\ee
\par
The solution to (\ref{8}) is the first-kind spherical Hankel function with argument $kr$ and order $\nu_{nm}$, where $\nu_{nm}(\nu_{nm}+1) = n^2 + m^2$.  Solving this quadratic equation, we find
\be{nu1}
\nu_{nm} = \frac{1}{2}\left[\sqrt{1+4(n^2+m^2)}-1\right]
\ee
to give\footnote{The order $\nu^\__{nm} \!=\! -(\sqrt{1+4(n^2+m^2)}+1)/2$ is also a solution to the quadratic equation.  But since there is just one linearly independent outgoing-wave solution to the second-order differential equation in (\ref{8}), $h_{\nu^\__{nm}}(kr)$ is not linearly independent of $h_{\nu_{nm}}(kr)$ and thus $h_{\nu^\__{nm}}(kr)$ need not be included.  This conclusion is confirmed by equation 10.1.18 of \cite{A&S1964}.}
\be{9}
B_{nm}(\w,r) = B_{nm}(\w)h_{\nu_{nm}}(kr)
\ee
which recasts (\ref{6}) as
\be{10}
E_{\w}(r,\theta,\phi) = \sum\limits_{m=-\infty}^{+\infty}\sum\limits_{n=-\infty}^{+\infty} B_{nm}(\w)h_{\nu_{nm}}(kr)e^{in\theta}e^{im\phi}
\ee
where the spherical mode coefficients $B_{nm}(\w)$ can be functions of the frequency $\w$.
This spherical mode expansion for the $\theta$ or $\phi$ polarization of the electric field has been derived for all values of $\phi$ but only for values of $\theta$ near $\pi/2$.  Specifically, the spherical mode expansion in (\ref{10}) applies with considerable accuracy within the free-space ($r\ge a$) biconical region defined by $[0\le\phi\le2\pi,\;\pi/2-\Delta\theta_0 < \theta < \pi/2+\Delta\theta_0]$ where $\Delta\theta_0$ is small enough that $\sin(\pi/2\!-\!\Delta\theta_0)\!=\!\cos\Delta\theta_0$ does not sub\-stan\-ti\-ally differ from $1$.  As far as the author is aware, the spherical expansion in (\ref{10}) has not been derived previously.
\par
There is no essential loss of generality for determining speckle patterns and correlation functions by restricting our observation angles for the spherical mode expansion to this $\Delta\theta_0$ angular region about $\pi/2$ since, as we will show next, the maximum range of $|m|$ and $|n|$ in the summation of (\ref{10}) is effectively $ka$ and thus the Fourier series in $\theta$ and $\phi$ produces an average angular beamwidth of the speckle pattern of a star on the order of $\lambda/D\; (D=2a)$, which is much narrower than the $\Delta\theta_0$ angular region, that is, $\lambda/D\ll\Delta\theta_0$.  With respect to Fig. \ref{Figstar}, the $\Delta\theta_0$ angular region is the conical wedge outside the star with the top and bottom surfaces of the conical wedge making an angle $\pm \Delta\theta_0$ with the $xy$ plane.
\par
Bounds on the modal coefficients $B_{nm}(\w)$ can be obtained by writing (\ref{10}) for $r=a$, that is, at the surface of the star in free space just outside the reactive fields
\bea{11}
E_{\w}(a,\theta,\phi) = \sum\limits_{m=-\infty}^{+\infty}\sum\limits_{n=-\infty}^{+\infty} B_{nm}(\w)h_{\nu_{nm}}(ka)e^{in\theta}e^{im\phi}.
\eea
For $\nu_{nm}> ka$, the magnitude of the spherical Hankel function $h_{\nu_{nm}}(ka)$ increases extremely rapidly and the electromagnetic fields associated with each Hankel function become extremely large and highly reactive.   However, the electric field $E_\w(a,\theta,\phi)$ is located in free space outside the reactive fields of the star.  These two facts are incompatible unless $B_{nm}(\w)\approx 0$ for $\nu_{nm}> ka$.\footnote{An accurate evaluation of the Hankel functions shows that the maximum value $\mathfrak{N}$ of $\nu_{nm}$ required for a relative error $\varepsilon$ in the field $E_\w(r,\theta,\phi)$ for $r\ge a$ is given by $\mathfrak{N} = ka[1+\frac{1}{2}(-3\ln\varepsilon/ka)^\frac{2}{3}]$  \cite[sec. 3.4.1]{Chewbook}, \cite{Rokhlin1995-in-Chews-book}, \cite{Bucci&Franceschetti-APS1987}.
Since $ka$ for visible light from a star with the typical diameter of the sun is on the order of $10^{16}$, $\mathfrak{N}$ equals $ka$ to an extremely high accuracy.}
With this bound on $B_{nm}(\w)$, (\ref{10}) can be rewritten as
\be{13}
E_{\w}(r,\theta,\phi) = \sum\hspace{-5mm}\sum\limits_{\hspace{-3mm}m,n (\nu_{nm}\le ka)}\hspace{-3mm}B_{nm}(\w)h_{\nu_{nm}}(kr)e^{in\theta}e^{im\phi}.
\ee
Changing the summation indices $m$ and $n$ to $u$ and $v$ in accordance with
\be{14}
m = uk a,\;\;n=vk a,\;\;{\sf w}=\sqrt{u^2+v^2}
\ee
(\ref{13}) becomes\\[-3mm]
\be{15}
E_{\w}(r,\theta,\phi) = \sum\hspace{-1mm}\sum\limits_{\hspace{-5mm}u,v ({\sf w}\le 1)}B_\w(u,v)h_{{\sf w}ka}(kr)e^{ivka\theta}e^{iuka\phi}\\[-2mm]
\ee
where $\nu_{nm}$ has been approximated as $\sqrt{n^2+m^2}$.  Because $ka$ for a star is so enormous, the increments $1/(ka)$ in $u$ and $v$ are extremely tiny and this approximation has a negligible effect on the value of $E_{\w}(r,\theta,\phi)$.
\par
The value of the far electric field of the star can be found by using the large-argument approximation for the spherical Hankel function in (\ref{15}), namely\\[-3mm]
\be{16}
h_{{\sf w}ka}(kr) \sim (-i)^{{\sf w}ka+1}\frac{e^{ikr}}{kr}\\[-2mm]
\ee
to get\\[-2mm]
\be{17}
E_{\w}(r,\!\theta,\!\phi)\! \sim\! \frac{e^{ikr}}{ikr}\!\sum\hspace{-1mm}\sum\limits_{\hspace{-5mm}u,v ({\sf w}\le 1)}\!\!\!B_\w(u,v)(-i)^{{\sf w}ka}e^{ivka\theta}e^{iuka\phi}.\\[-1mm]
\ee
In the visible spectrum, the Earth is not in the single-frequency (monochromatic) far field of the sun and some other stars ($r\not>2a^2/\lambda$).  In fact, the Earth is not even in the sun's single-frequency Fresnel zone, which begins a distance equal to about $a\sqrt{2a/\lambda}$ from the sun. However, the more accurate large-$r$ approximation to the Hankel function leads to $e^{ikr\{1+O[({\sf w}a/r)^2]\}}/(kr)$ instead of $e^{ikr}/(kr)$ and it can be shown, as was done in Section \ref{sub1:Quasi} for the line source, that the $O[({\sf w}a/r)^2]$ terms have negligible effect on the statistical character of the quasi-monochromatic fields as long as 
\be{crit'}
r\ga 2\,{\rm Min}\!\left[a\sqrt{\w_0/\Delta\w_0}\,,a^2/\lambda_0\right]
\ee
where $\w_0$ is the mean angular frequency and $\Delta\w_0$ is the bandwidth of the measurement system.  
\par
The phases of the $\pi(ka)^2$ number of modal coefficients $B_\w(u,v)$ in (\ref{15}) are randomly distributed with equal probability between $0$ and $2\pi$ for different values of $(u,v)$  because they are linearly generated by the many more randomly phased volume sources within the star.  (This randomness of the $B_\w(u,v)$  has been confirmed by numerical computations.)  Therefore, since the Hankel functions in (\ref{15}) depend only on the $r$ coordinate but not on the $\theta$ or $\phi$ coordinates, the statistical character of the $\theta$ and $\phi$ variation of the fields for a fixed $r$ does not change for all $r$ in (\ref{15}).  That is, the far-field approximation in (\ref{17}) can be used for determining the angular field behavior and angular correlation functions at a fixed $r$ everywhere outside the reactive fields of the star ($r>a$).  This radial independence of the angular variation of the fields of spherical stars was also predicted from the results obtained for line sources in Section \ref{sub1:Quasi} and from the numerical computations of the cross spectral density in \cite{AGW} for a spherical model of the sun. 
\subsection{Quasi-Monochromatic Fields of the Star}\lbl{subsec:Quasi}
The expressions developed so far have been for single-frequency (continuous-wave, monochromatic) fields.  However, most stellar interferometry measurements involve narrow-band (quasi-monochromatic) time-domain fields, such as the  fields within a portion of the visible spectrum.  The narrow bandwidths can be an inherent restriction of the photoelectric detectors or they can be produced by filtering at some other stage of the measurement process.  These time-domain quasi-monochromatic fields can be found by taking the inverse Fourier transform of the frequency-domain fields.  Specifically, for the electric field in spherical coordinates, we have\\[-5mm]
\be{18}
E(r,\theta,\phi,t)\!=\!\!\!\int\limits_{-\infty}^{+\infty}\!\!\!E_{\w}(r,\!\theta,\!\phi)e^{-i\w t}\! d\w = 2\mbox{Re}\!\!\int\limits_0^{\infty}\!\!E_{\w}(r,\!\theta,\!\phi)e^{-i\w t} \!d\w
\ee
where the second equality in (\ref{18}) follows from $E(r,\theta,\phi,t)$ being a real function.  Twice the Fourier transform over positive frequencies only in (\ref{18}), before taking the real part, gives a complex electric field that is referred to as the analytic-signal electric field \cite[sec. 5.3]{H&Y}.\footnote{We will not work directly with the complex analytic-signal fields in this paper, as is commonly done for partially coherent fields \cite{Wolf},\cite{B&P}.  The alternative approach used here has the advantage of dealing with the actual real time-domain fields and with real correlation functions that are measured with stellar interferometers.  Using real time-domain fields proves especially convenient for determining the second-order correlation function measured in Hanbury Brown--Twiss intensity stellar interferometry and the simple  relationship of this second-order correlation function to the first-order correlation function measured in Michelson phase stellar interferometry; see Section \ref{sec:Second}.}
Applying (\ref{18}) to the far electric field in (\ref{17}) gives\\[-5mm]
\bea{19}
E(r,\pi/2+\Delta\theta,\phi,t)\hspace{45mm}\nonumber\\ \sim \frac{2}{r}\mbox{Re}\int\limits_0^{\infty} \sum\hspace{-1mm}\sum\limits_{\hspace{-5mm}u,v ({\sf w}\le 1)} b_\w(u,v) e^{i\frac{\w}{c}(r +va\Delta\theta+ua\phi -ct)} d\w
\eea
\mbox{}\\[-4mm]
with
\be{20}
b_\w(u,v) = B_\w(u,v)(-i)^{({\sf w}-v)ka}/ik
\ee
where the substitution $\theta = \pi/2+\Delta\theta$ has been made to indicate that this expression holds for $\Delta\theta \le \Delta\theta_0$ such that $\cos{\Delta\theta_0} \approx 1$.  For example, if $\cos{\Delta\theta_0}=0.99$, then $\Delta\theta_0 = 8.11$ degrees.
\par
For a bandlimited signal whose spectrum falls off rapidly be\-yond a bandwidth $\pm\Delta\w_0/2$ from a center frequency $\w_0$, so that the effective bandwidth is $\Delta\w_0$ and the fractional band\-width is much less than unity  ($\Delta\w_0/\w_0\!\! \ll\!\! 1$), we can change the integration variable $\w$ to $\Delta\w \!=\! \w\!-\!\w_0$ and rewrite (\ref{19}) as
\bea{21}
&&\hspace{-7mm}E(r,\pi/2+\Delta\theta,\phi,t) \sim \frac{2}{r}\mbox{Re}\sum\hspace{-1mm}\sum\limits_{\hspace{-5mm}u,v ({\sf w}\le 1)}e^{i\frac{\w_0}{c}(r +va\Delta\theta+ua\phi -ct)}\nonumber\\[-3mm]&&\cdot\int\limits_{-\Delta\w_0/2}^{+\Delta\w_0/2}b_{\w_0+\Delta\w}(u,v)e^{i\frac{\Delta\w}{c}(r +va\Delta\theta+ua\phi -ct)} d\Delta\w.
\eea
Since $\Delta\w_0/\w_0 \ll 1$\\[-3mm]
\be{22}
u = m/ka \approx m/k_0a,\;\;v = n/ka \approx m/k_0a\\[-2mm]
\ee
where $k_0 =\w_0/c$, and it is permissible to interchange the integration and summations in (\ref{19}) to obtain (\ref{21}).  Defining the time-domain complex spherical mode coefficients by\\[-5mm]
\bea{23}
b(u,v,t) =\!\!\! \int\limits_{-\Delta\w_0/2}^{+\Delta\w_0/2}b_{\w_0+\Delta\w}(u,v)e^{-i\Delta\w t} d\Delta\w
\eea
\mbox{}\\[-4mm]
transforms (\ref{21}) to\\[-4mm] 
\bea{24}
E(r,\pi/2,\phi,t) \sim \frac{2}{r}\sum\hspace{-1mm}\sum\limits_{\hspace{-5mm}u,v ({\sf w}\le 1)}\mbox{Re}\Big\{b[u,v,t\!-\!(r+ua\phi)/c]\nonumber\\[-3mm]\cdot e^{i\frac{\w_0}{c}(r +ua\phi -ct)}\Big\}
\eea
\mbox{}\\[-5mm]
where we have chosen $\Delta\theta =0$ to shorten the expression and subsequent derivation somewhat by concentrating on the variation in $\phi$ at $\theta=\pi/2$.  The spherical symmetry of the star demands that the average angular speckle size and cross correlation functions are the same in the $\theta$ and $\phi$ directions.
Because $\Delta\w_0/\w_0 \ll 1$, the complex time dependent function $b(u,v,t)$ has a minimum time period equal to $2\pi/\Delta\w_0$, a period that is much longer than the period $2\pi/\w_0$ of the center frequency.  Thus, the time dependence of each spherical mode field as well as the time-domain field $E(r,\theta,\phi,t)$ is a slowly modulated sinusoidal wave with carrier frequency $\w_0$ and an average modulation frequency approximately equal to $\Delta\w_0$, as was also seen for the line source in Section \ref{sec:TCM}.
\par
Expressing the complex time dependent function $b(u,v,t)$ in terms of its magnitude and phase\\[-3mm]
\be{25}
b(u,v,t) = |b(u,v,t)|e^{-i\psi(u,v,t)}\\[-2mm]
\ee
allows (\ref{24}) to take the form of a sum of ``quasi-monochromatic'' cosine waves\\[-4mm]
\bea{26}
E(r,\pi/2,\phi,t) \sim \frac{2}{r}\sum\limits_{(uv)}\!|b(u,v,\tau)|\cos{[\w_0\tau+\psi(u,v,\tau)]}
\eea
\mbox{}\\[-3mm]
with the shorter notation $\sum\limits_{(uv)}\! =\! \sum\hspace{-1mm}\sum\limits_{\hspace{-4mm}u,v ({\sf w}\le 1)}$
and $\tau\! = \!t\!-\!(r \!+\!ua\phi)/c$.  Because $\Delta\w_0/\w_0 \!\ll\! 1$, the phase functions $\cos[\psi(u,v,\tau)]$ and the magnitude functions $|b(u,v,\tau)|$ vary much more slowly with $\tau$ than $\cos\w_0\tau$.  The quasi-monochromatic fields form a train of wave packets propagating in the radial direction at each $(\theta,\phi)$; see Fig. \ref{figwavetrain}.  The minimum radial wave-packet length $\Delta r$ is approximately
\be{28'}
\Delta r = 2\pi c/\Delta\w_0 = (\w_0/\Delta\w_0)\lambda_0
\ee
which is much larger than the wavelength $2\pi c/\w_0 = \lambda_0$ of the center (carrier) frequency.  Note that the minimum length $\Delta r$ of the wave packets is inversely proportional to the bandwidth.
\section{First-Order Correlation Functions}\lbl{sec:First}
The $E(r,\pi/2,\phi,t)$ in (\ref{26}) represents the $\theta$ or $\phi$ polarized electric field of the star at the space-time point $(r,\pi/2,\phi,t)$ under the criterion given in (\ref{crit'}).  To get a measure of the partial coherence of the fields at two different values of $\phi$, we want to determine the first-order temporally averaged correlation function $G^{(1)}(\Delta\phi)$ of either polarization of the electric field for the two space-time points $(r,\pi/2,\phi_1,t)$ and $(r,\pi/2,\phi_2,t)$, where $\Delta\phi = \phi_2-\phi_1$; specifically\\[-3mm]
\be{29}
G^{(1)}(\Delta\phi) = \langle E(r,\pi/2,\phi_1,t)E(r,\pi/2,\phi_2,t)\rangle\\[-1mm]
\ee
where the brackets $\langle\;\rangle$ denote the time (temporal) average defined for a quasi-monochromatic function of time $f(t)$ beginning at time $t=0$ as in (\ref{P6}).
A Michelson phase stellar interferometer essentially measures the correlation function in (\ref{29}).   In practice, the time $T$ used in the averaging in (\ref{P6}) has a finite value $T\gg 2\pi/\Delta\w_0$.
\par
Insertion of the electric field from (\ref{26}) into (\ref{29}) gives\\[-4mm]
\bea{31}
G^{(1)}(\Delta\phi) = \frac{4}{r^2}\sum\limits_{(uv)}\sum\limits_{(u'v')}\langle|b(u,v,\tau_1)b(u',v',\tau'_2)|\nonumber\hspace{8mm}\\[-1mm]\cdot \cos{[\w_0\tau_1+\psi(u,v,\tau_1)]}\cos{[\w_0\tau'_2+\psi(u',v',\tau'_2)]}\rangle
\eea
with\\[-4mm]
\be{32}
\tau_1 = t-(r +ua\phi_1)/c,\;\;\tau'_2 = t-(r +u'a\phi_2)/c.
\ee
The product of the cosines in (\ref{31}) can be rewritten as\\[-5mm]
\bea{33}
\{\cos[\w_0(\tau_1-\tau'_2) +\psi(u,v,\tau_1)-\psi(u',v',\tau'_2)]\nonumber\hspace{3mm}\\
+\cos[\w_0(\tau_1+\tau'_2) +\psi(u,v,\tau_1)+\psi(u',v',\tau'_2)]\}/2.
\eea
\mbox{}\\[-5mm]
The last cosine term in (\ref{33}) oscillates at approximately the rate $\cos{2\w_0t}$ and, thus, contributes a negligible amount to the time average so that (\ref{31}) reduces to\\[-4.5mm]
\bea{34}
G^{(1)}(\Delta\phi) = \frac{2}{r^2}\sum\limits_{(uv)}\sum\limits_{(u'v')}\langle|b(u,v,\tau_1)b(u',v',\tau'_2)|\nonumber\\[-1mm] \cdot\cos[\w_0(\tau_1-\tau'_2) +\psi(u,v,\tau_1)-\psi(u',v',\tau'_2)]\rangle.
\eea
\mbox{}\\[-5.5mm]
The different coefficients $b(u,v,t)$ of the $\pi(k_0a)^2$ spherical modes are uncorrelated, that is, the $b(u,v,t)$ have zero temporal correlation for different values of $(u,v)$, because the volume sources of stellar radiation are statistically independent and number many more than the number of required modal coefficients.  (This zero correlation has been confirmed by numerical computations.)   Therefore, except for $(u',v') = (u,v)$, the phase function $\psi(u,v,\tau_1)-\psi(u',v',\tau'_2)$ varies with time $t$ from $0$ to $2\pi$ at an average frequency on the order of $\Delta\w_0$.  Consequently, after averaging for a time $T\gg 2\pi/\Delta\w_0$, all the terms in the two-fold double summation of (\ref{34}) become negligible  except for $(u',v') = (u,v)$ and (\ref{34}) reduces to the one-fold double summation\\[-4mm]
\bea{35}
G^{(1)}(\Delta\phi) = \frac{2}{r^2}\sum\limits_{(uv)}\langle|b(u,v,\tau_1)b(u,v,\tau_2)|\nonumber\hspace{10mm}\\[-1mm] \cdot\cos[\w_0(\tau_1-\tau_2) +\psi(u,v,\tau_1)-\psi(u,v,\tau_2)]\rangle.
\eea
\mbox{}\\[-5mm]
The correlation function $G^{(1)}(\Delta\phi)$ is significant only for values of $(\tau_1\!-\!\tau_2)\! \la\! 2\pi/\w_0$, whereas the phase function  $\cos[\psi(u,v,\tau_1)\!-\!\psi(u,v,\tau_2)]$ varies very little with $(\tau_1\!-\!\tau_2)$ over this range of values because $\cos[\psi(u,v,\tau)]$ varies with $\tau$ at an average frequency on the order of $\Delta\w_0\! <<\!\w_0$.   Thus, we can set $\psi(u,v,\tau_1)\!\approx\!\psi(u,v,\tau_2)$.  For the same reason, we can set $|b(u,v,\tau_2)| \!\approx \!|b(u,v,\tau_1)|$ to further reduce (\ref{35}) to\\[-4mm]
\bea{36}
G^{(1)}(\Delta\phi) = \frac{2}{r^2}\sum\limits_{(uv)}\langle|b(u,v,t)|^2\rangle \cos{(k_0a\Delta\phi u)}
\eea
\mbox{}\\[-3mm]
where we have used (\ref{32}) to write $\w_0(\tau_1-\tau_2)= k_0a(\phi_2-\phi_1)u = k_0a\Delta\phi u$.  Also, since the correlation function does not depend on the observation angle $\phi_1$, the time average of the intensity of the spherical mode coefficients cannot change with observation angle and thus we have written $\langle|b(u,v,t)|^2\rangle=\langle|b(u,v,\tau_1)|^2\rangle$ in (\ref{36}).
\subsection{Spherical Mode Coefficients for a Lambertian Star}\label{subsec:Lambertian}
To evaluate the first-order correlation function in (\ref{36}), we have to know the functional dependence with $(u,v)$ of the time-average of the intensity of the time-domain spherical mode coefficients, that is, $\langle|b(u,v,t)|^2\rangle$.  This functional dependence can be found from the assumption that within the given quasi-monochromatic bandwidth the star radiates as a spherical Lambertian source.
In order to quantify Lambertian radiation, consider an optical antenna (for example, a telescope with a hypothetical single photoelectric detector confined to its central focal spot) with a given resolution angle (beamwidth).  This telescope is located a large distance $R_0$ from the star and is directed toward the point 
$(r=a, \theta=\pi/2, \phi=0)$ on the surface of the star.  In terms of the $xyz$ coordinate system shown in Fig. \ref{Figstar}, the coordinates of this point on the star are $(x=a,y=0,z=0)$.  Imagine the telescope remains directed toward this point at a large constant distance $R_0$ from the point but can be positioned in space to make an angle $\alpha$ with the $x$ axis, where $0\le\alpha<\pi/2$. Furthermore, assume that the telescope has a narrow enough resolution angle that it subtends only the volume sources (molecules within a depth $L_0$ along the direction of $\alpha$) beneath a small area $A_\alpha=A_0/\cos\alpha$ at the surface of the star, where $A_0$ is the area subtended when the telescope is directed normal to the surface ($\alpha=0$).\footnote{The spherical wave emanating from each molecular source within $A_{\alpha}$ has virtually uniform phase across the aperture of the telescope and thus behaves as a plane wave incident on the telescope from the direction of the molecular source to the telescope.  For directions within the resolution beamwidth of the telescope, the fields in the plane waves will be focused (for the most part) and received on the photoelectric detector.  For directions outside this beamwidth, the plane-wave fields will not be focused (for the most part) on the photoelectric detector and not be received.  Alternatively, one can assume a fixed telescope directed along $\alpha=0$ but with many detectors in the focal plane (similar to a photographic film), each one receiving the power from successive constant resolution angles across the diameter of the star.}  Assume the resolution of the telescope is so narrow that the area $A_\alpha$ on the surface of the star is small enough to be practically planar while remaining electrically large enough ($k\sqrt{A_0}\gg 1$) to cover a vast number of molecular sources.  
\par
If the surface of the star is Lambertian, the intensity of the radiation received by the telescope is independent of the observation angle $\alpha$ for each polarization.  In terms of the $\theta$ or $\phi$ component of the far electric fields (with the $1/r$ dependence removed), denoted by $F_{A_\alpha}(\alpha,t)$ and radiated by the volume sources a depth $L_0$ beneath the surface area $A_\alpha$, we have $\langle|F_{A_\alpha}(\alpha,t)|^2\rangle = F_0^2$, where $F_0$ is a constant independent of $\alpha$.  The subscript $A_\alpha$ on $F_{A_\alpha}$ indicates that it is the far field radiated by sources in an area that increases with $\alpha$ as $1/\cos\alpha$.  Since the volume sources that radiate this far field are statistically independent, the far field intensity is proportional to the number of sources.  Consequently, the far-field intensity $\langle|F(\alpha,t)|^2\rangle$ radiated by the sources within a fixed area $A$ on the surface of the star is given by\\[-4mm] 
\be{37}
\left\langle|F(\alpha,t)|^2\right\rangle = F_0^2 \cos{\alpha}\\[-2mm]
\ee
which is commonly referred to as Lambert's law \cite[sec. 5.1]{Wolf}.  
\par
Though we assume the stars are Lambertian radiators, limb effects in terms of a known symmetric in\-ten\-sity taper can be taken into account by making $F_0^2$ a function of $\alpha$ in (\ref{37}) and the succeeding equations.  This would change the integrals in (\ref{52})--(\ref{53}) and (\ref{89})--(\ref{92}) to give $g^{(1)}(\Delta\phi)$ and $g^{(2)}(\Delta\phi)$ modified from those in (\ref{55}) and (\ref{97}) according to the particular limb darkening/brightening (intensity taper).
\par
The far field intensity in (\ref{37}) can be related to the intensity of the time-domain spherical mode coefficients $\langle|b(u,v,t)|^2\rangle$ in (\ref{36}) by starting with the expression in (\ref{15}) for the frequency-domain field at the surface of the star ($r=a$) and $\theta=\pi/2$, namely\\[-3mm] 
\be{38}
E_{\w}(a,\pi/2,\phi) = \sum\limits_{(uv)}B_\w(u,v)h_{{\sf w}ka}(ka)(i)^{vka}e^{iuka\phi}.\\[-3mm]
\ee
Approximating the double summation in (\ref{38}) by a double integral and using (\ref{20}), we have\\[-4mm]
\be{40}
E_{\w}(a,\!\pi/2,\!\phi)\!\! =\!\! ik(ka)^2\!\!\!\!\!\!\iint\limits_{u,v({\sf w}\le 1)}\!\!\!\!\!\!b_\w(u,v)h_{{\sf w}ka}(ka)(i)^{{\sf w}ka}e^{iuka\phi}dudv\\[-.5mm]
\ee
which is the propagating plane-wave representation with respect to the coordinate $a\phi$ for the frequency-domain electric field at the surface of the star where\\[-3mm]
\be{41}
 T_\w(u,v)= \frac{ik(ka)^2}{(2\pi)^2}b_\w(u,v)h_{{\sf w}ka}(ka)(i)^{{\sf w}ka}\\[-2mm]
\ee
is the propagating plane-wave spectrum  \cite[sec. 3.2]{H&Y}.  The propagating plane-wave spectrum $T_\w(u,v)$ can be expressed in terms of the far electric field $F_\w(u,v)$ of the sources producing $E_{\w}(a,\pi/2,\phi)$ as \cite[eq. (3.112)]{H&Y}\\[-3mm]
\be{42}
T_\w(u,v) = \frac{iF_\w(u,v)}{k\sqrt{1-{\sf w}^2}}\\[-2mm]
\ee
so that (\ref{41}) and (\ref{42}) reveal that\\[-3mm]
\be{43}
b_\w(u,v) =  \frac{(2\pi)^2}{k^2(ka)^2}\frac{F_\w(u,v)}{h_{{\sf w}ka}(ka)\sqrt{(1-{\sf w}^2)}}.\\[-2mm]
\ee
The magnitude squared of (\ref{43}) gives\\[-3mm]
\be{44}
|b_\w(u,v)|^2 =  \frac{(2\pi)^4}{k^4(ka)^4}\frac{|F_\w(u,v)|^2}{|h_{{\sf w}ka}(ka)|^2(1-{\sf w}^2)}.\\[-1mm]
\ee
Since ${\sf w}\le 1$, it follows that ${\sf w}ka \le ka$ and an accurate uniform asymptotic expansion of $|h_{{\sf w}ka}(ka)|^2$ for all ${\sf w}\le 1$ and $ka \gg 1$ is given by \cite[sec. 4.3, eq. (7)]{Jeffreys}, \cite[eq. (39)]{YaghjianCylTN}\\[-3mm]
\be{45}
|h_{{\sf w}ka}(ka)|^2 = \frac{1}{(ka)^2\sqrt{(1 -{\sf w}^2)}}\\[-2mm]
\ee
which simplifies (\ref{44}) to\\[-3mm]
\be{46}
|b_\w(u,v)|^2 =  \frac{(2\pi)^4}{k_0^4(k_0a)^2}\frac{|F_\w(u,v)|^2}{\sqrt{(1-{\sf w}^2)}}\\[-2mm]
\ee
where we have replaced $k$ by $k_0$ for narrow-band (quasi-monochromatic) fields.  
\par
Integrating (\ref{46}) over all frequencies and invoking Parceval's theorem for Fourier transforms reveal that the relationship between the squares of the frequency-domain spectra in (\ref{46}) holds also for the long-time averages of the squares of the time-domain spectra, that is\\[-3mm] 
\be{47}
\langle|b(u,v,t)|^2\rangle =  \frac{(2\pi)^4}{k_0^4(k_0a)^2}\frac{\langle|F(u,v,t)|^2\rangle}{\sqrt{(1-{\sf w}^2)}}.\\[-2mm]
\ee
The time-domain far field $F(u,v,t)$ is that of the sources that produce the electric field $E(a,\pi/2,\phi,t)$ at the surface of the star.  These sources lie within a finite radius ($L_0$) of the surface point $(a,\pi/2,\phi)$.  Thus, Lambertian's law in (\ref{37}) applies to $F(u,v,t)$.   Noting that in the $(u,v)$ plane-wave representation\\[-3mm]
\be{49}
\langle|F(\alpha,t)|^2\rangle = \langle|F(u,v,t)|^2\rangle,\;\;
\cos{\alpha} = \sqrt{1-{\sf w}^2}\\[-1mm]
\ee
we obtain from (\ref{37}) and (\ref{47}) the result
\be{50}
\langle|b(u,v,t)|^2\rangle =  \frac{(2\pi)^4 F_0^2}{k_0^4(k_0a)^2}.
\ee
In other words, \textit{for a Lambertian spherical star, the time-average of the intensity of the time-domain spherical mode coefficients equals the same constant for all the mode numbers $(u,v)$ in the propagating spectrum defined by ${\sf w}^2 \!=\! u^2+v^2\! \le\! 1$}.
\par
Summing (\ref{50}) over $(uv)$ gives\\[-3mm]
\be{50sum1}
\sum\limits_{(uv)}\langle|b(u,v,t)|^2\rangle =  \Big\langle\sum\limits_{(uv)}|b(u,v,t)|^2\Big\rangle = \frac{(2\pi)^4\pi F_0^2}{k_0^4}\\[-2mm]
\ee
since there are $\pi(k_0a)^2$ values of $(uv)$ within the circle ${\sf w}^2 \le 1$.
For later reference, we denote the constant in (\ref{50sum1}) as $b_0^2$
\be{50sum2}
b^2_0 = (2\pi)^4\pi F_0^2/k_0^4.
\ee
\subsection{Evaluation of the First-Order Temporally Averaged Correlation Function}\label{subsec:Evaluation Temporally}
With $\langle|b(u,v,t)|^2\rangle$ from (\ref{50}) substituted into (\ref{36}), it becomes relatively easy to evaluate $G^{(1)}(\Delta\phi)$
\be{51}
G^{(1)}(\Delta\phi) = \frac{2(2\pi)^4 F_0^2}{k_0^4(k_0a)^2 r^2}\sum\limits_{(uv)} \cos{(k_0a\Delta\phi u)}
\ee
by approximating the summations with the integrals
\be{52}
G^{(1)}(\Delta\phi) = \frac{2(2\pi)^4 F_0^2}{k_0^4 r^2}\!\!\!\!\iint\limits_{u,v({\sf w}\le 1)}\!\!\! \cos{(k_0a\Delta\phi u)}dv du.
\ee
Since ${\sf w}^2 \le 1$, the integration in (\ref{52}) can be expressed as\\[-5mm]
\bea{53}
G^{(1)}(\Delta\phi) = \frac{2(2\pi)^4 F_0^2}{k_0^4 r^2} \int\limits_{-1}^{+1} \cos{(k_0a\Delta\phi u)}\bigg(\int\limits_{-\sqrt{(1-u^2)}}^{+\sqrt{(1-u^2)}}\!\!\!\!\!\!dv\bigg) du\nonumber\hspace{-5mm}\\[-3mm]
= \frac{4(2\pi)^4 F_0^2}{k_0^4 r^2} \int\limits_{-1}^{+1}\sqrt{(1-u^2)} \cos{(k_0a\Delta\phi u)} du.\hspace{5mm}
\eea
\mbox{}\\[-4mm]
With the help of \cite[3.752]{G&R}, the integral in (\ref{53}) evaluates to
\be{54}
G^{(1)}(\Delta\phi) = \frac{4\pi(2\pi)^4 F_0^2}{k_0^4 r^2} \frac{J_1(k_0a\Delta\phi)}{k_0a\Delta\phi}
\ee
where $J_1(x)$ is the first-order Bessel function.
\par
The normalized first-order temporally averaged correlation function $g^{(1)}(\Delta\phi)$ is defined as
\be{55}
g^{(1)}(\Delta\phi)= \frac{G^{(1)}(\Delta\phi)}{G^{(1)}(0)} = 2\frac{J_1(k_0a\Delta\phi)}{k_0a\Delta\phi}= 4\frac{J_1(k_0D\Delta\phi/2)}{k_0D\Delta\phi}.
\ee
It is independent of the angle $\phi$ and with $\sin\Delta\phi$ replacing $\Delta\phi$ in (\ref{55}), it is identical to the normalized first-order temporally averaged correlation function obtained near the normal to a circular disc model for a spherical star with incoherent, uniform-intensity sources \cite[sec. 3.3.1]{Wolf}, \cite[sec. 5.4]{B&P}.  Also with $\sin\Delta\phi$ replacing $\Delta\phi$, it is identical to the far-field pattern of a circular-aperture, monochromatic (fully coherent) field with uniform phase and magnitude \cite[sec. 10.5]{Johnson}.
\par
The first zero of $J_1(x)$ occurs at $x = 3.83$ so that the first-order correlation function  has its first zero at the angle
\be{56}
\Delta\phi_0 = 1.22 \frac{\lambda_0}{D}
\ee
the classic equation used by Michelson and Pease \cite{Michelson}, \cite{Pease} for their phase stellar interferometric determination of the diameters of stars.  (Apparently, Michelson \cite{Michelson} simply used a uniform phase and magnitude circular aperture to obtain (\ref{55}) and (\ref{56}).)  The zero-correlation width $d_0$ corresponding to the zero-correlation angle in (\ref{56}) for the light of the star a radial distance $R_0$ from the star is given by
\be{57}
d_0 \approx  R_0 \Delta\phi_0 = 1.22 \lambda_0\frac{R_0}{D}.
\ee
For our sun ($D= 1.39\times 10^9$ meters), the zero-correlation width on the Earth ($R_0= 1.49\times 10^{11}$ meters) is\\[-3mm]
\be{58}
d_0 \approx  130.5\times \lambda_0\\[-2mm] 
\ee
or with $\lambda_0 = 550$ nanometers, that is, the center wavelength of visible light, the zero-correlation width is\\[-3mm]
\be{59}
d_0 \approx  .072 \quad \mbox{millimeters}\\[-2mm]
\ee
which is too small to be measured by Michelson interferometry.  This correlation width for visible light from the sun
implies that computations of rainbow patterns of millimeter size raindrops using the Mie solution may require that the
plane waves emitted from sources across the sun's diameter be taken into account to obtain accurate results \cite{Shore}.  The ratios of $R_0/D$ for other stars are much larger than that of the sun and, thus, the zero-correlation widths are proportionally much larger.  For the star Betelgeuse, whose $R_0/D$ was determined by Pease from measurements with the first Michelson stellar interferometer \cite{Michelson}, \cite{Pease}\\[-3.5mm]
\be{60}
d_0 \approx  3.31 \quad \mbox{meters}.\\[-1.5mm] 
\ee
Next it will be proven that the normalized, angular, first-order, \textit{spatially} averaged correlation is identical to the normalized, first-order, temporally averaged correlation in (\ref{55}).
\subsection{Evaluation of the First-Order Spatially Averaged Correlation Function}\label{subsec:Evaluation Spatially}
Suppose that we could instantaneously (that is, at any one time $t$) measure the $\theta$ or $\phi$ component of the far electric fields of the star and ask what the speckle pattern would look like, say, as a function of $\phi$ at a fixed radius $r$.  Specifically, we can evaluate the first-order spatially averaged correlation function between two values of $\phi$ separated by $\Delta\phi$, namely
\be{61}
G_s^{(1)}(\Delta\phi) = \langle E(r,\pi/2,\phi,t)E(r,\pi/2,\phi + \Delta\phi,t)\rangle_s
\ee
where the brackets $\langle\;\rangle_s$ denote the spatial average over $\phi$ defined for a function $f(\phi)$ as\\[-3mm]
\be{62}
\langle f(\phi) \rangle_s = \frac{1}{2\pi}\int\limits_0^{2\pi} f(\phi) d\phi.\\[-2mm]
\ee
Insertion of the electric fields from (\ref{26}) into (\ref{61}) shows that the result contains integrations of the form\\[-3mm]
\be{63}
\frac{1}{2\pi}\int\limits_0^{2\pi} f[t-(r+ua\phi)/c] d\phi = \frac{1}{T}\int\limits_0^{T}f(t)dt\\[-2mm]
\ee
where $T \!=\! 2\pi ua/c$. For $T\! \gg \!2\pi/\Delta\w_0$ or, equivalently, $\Delta\w_0 a/c\! \gg \!1/u$, which holds for the practical bandwidth mea\-sure\-ments of all stars at optical wavelengths except for an extremely small range of values of $u$ near zero that contributes negligibly to the fields.  Thus, (\ref{63}) shows that the spatial average $\langle\;\rangle_s$ is equivalent to the time average $\langle\;\rangle$ and this implies that the first-order spatially averaged correlation $G_s^{(1)}(\Delta\phi)$ in {(\ref{61}) is equal to the first-order temporally averaged correlation $G^{(1)}(\Delta\phi)$ in (\ref{29}), that is\\[-3mm]
\be{64}
G_s^{(1)}(\Delta\phi) = G^{(1)}(\Delta\phi).\\[-2mm]  
\ee
Thus, the normalized first-order spatially averaged correlation function, $g_s^{(1)}(\Delta\phi) = G_s^{(1)}(\Delta\phi)/G_s^{(1)}(0)$, is equal to the normalized first-order temporally averaged correlation function\\[-3mm]
\be{65}
g_s^{(1)}(\Delta\phi)= g^{(1)}(\Delta\phi).\\[-1mm]
\ee
Since the first zero of $g_s^{(1)}(\Delta\phi)$ is approximately equal to the average angular half-power beamwidth of the lobes in the $\theta$ or $\phi$ component of the electric-field intensity pattern of the star at any instant of time $t$ at a fixed radius $r$, this average angular speckle beamwidth $\Delta\phi_s$ is given from (\ref{65}) and (\ref{56}) as\\[-3.5mm]
\be{66}
\Delta\phi_s = 1.22 \frac{\lambda_0}{D}.\\[-1mm]
\ee
A typical speckle pattern versus $\phi$ for the spherical star is just like the one given in Fig. \ref{figfieldpattern} for a linear array of incoherent sources except for the linear array having  $\Delta\phi_s = \lambda_0/D$.
\par
The corresponding average half-power linear width $d_s$ of the speckles at a radial distance $R_0$ from the star is given by\\[-2mm]
\be{66'}
d_s = 1.22 \lambda_0\frac{R_0}{D}.\\[-1mm]
\ee
Note that, unlike the minimum length $\Delta r$ given in (\ref{28'}) for the wave packets, the average width of the wave packets, as exhibited by the average half-power linear speckle width ($d_0=d_s$), does not depend upon the bandwidth (for small fractional bandwidths) but only on the wavelength $\lambda_0$, that is, the center frequency $\w_0$, for a given $R_0/D$.  The average angular width of narrow-band wave packets depends on the center frequency and the diameter of the star, but not on the bandwidth, whereas the minimum length of the wave packets depends only on the bandwidth of the detector/receiver.
\section{Second-Order Temporally Averaged Correlation Function}\lbl{sec:Second}
The first-order temporally averaged correlation function in (\ref{29}) requires that the relative phase (time delay) between $E(r,\pi/2,\phi_1,t)$ and $E(r,\pi/2,\phi_2,t)$ be preserved as the time average of their product is measured with a Michelson stellar interferometer.  High tolerances required by Michelson stellar interferometry led to the development by Hanbury Brown--Twiss in the 1950's of an ``intensity interferometer'' \cite{HBT1956}, \cite{HBT1957}, which requires only the magnitudes  $E^2(r,\pi/2,\phi_1,t)$ and $E^2(r,\pi/2,\phi_2,t)$ measured by photoelectric detectors to be temporally correlated to obtain the diameters of stars.  The key to the success of the Hanbury Brown--Twiss interferometry is that the photoelectric detectors have a response time much longer than the center-frequency period but much shorter than the minimum modulation period of the measured quasi-monochromatic fields.   In award winning writer Marcia Bartusiak's brief history of optical interferometry through 1982 \cite{MB}, she says that, ``According to his [Hanbury Brown's] chronicle of the project, the function of each of its [the intensity-interferometer's] 6.5-meter telescopes was to collect the light from the star, `like rain in a bucket, and pour it into a [photoelectric] detector'.'' 
\par
Consider a photoelectric detector measuring the light intensity (magnitude squared) of the $\theta$ or $\phi$ polarization of the incident electric field.  Assume that the response time $T_d$ of the photoelectric detector is much longer than the period $2\pi/\w_0$ of the mean frequency and much shorter than the minimum period $2\pi/\Delta\w_0$ of the modulation frequency, that is\\[-2mm]
\be{67}
2\pi/\w_0 \ll T_d \ll 2\pi/\Delta\w_0\\[-1mm]
\ee
which implicitly requires $\Delta\w_0/\w_0 \ll 1$.  Placing the first photoelectric detector at the angle $\phi_1$, it measures the running average $\overline{E^2(r,\pi/2,\phi_1,t)}$, where the overline denotes the running average over the response time $T_d$ of the detector.   We have from (\ref{26})\\[-3mm]
\bea{68}
\overline{E^2(r,\pi/2,\phi_1,t)} = \frac{4}{r^2}\sum\limits_{(u_1v_1)}\sum\limits_{(u_1'v_1')}|b(u_1,v_1,\tau_1)b(u_1',v_1',\tau'_1)|\nonumber\hspace{-6mm}\\\cdot \overline{\cos{[\w_0\tau_1+\psi(u_1,v_1,\tau_1)]}\cos{[\w_0\tau'_1+\psi(u_1',v_1',\tau'_1)]}}\hspace{5mm}
\eea
with
\be{69}
\tau_1 = t-(r +u_1a\phi_1)/c,\;\;\tau'_1 = t-(r +u_1'a\phi_1)/c.
\ee
The time variation of $|b(u,v,t)|$ has a minimum period of about $2\pi/\Delta\w_0 \gg T_d$ so that $\overline{|b(u_1,v_1,\tau_1)b(u_1',v_1',\tau'_1)|}$ has been set equal to $|b(u_1,v_1,\tau_1)b(u_1',v_1',\tau'_1)|$ in (\ref{68}).
The product of the cosines in (\ref{68}) can be rewritten as
\bea{71}
\textstyle\frac{1}{2}\cos[\w_0(\tau_1-\tau'_1) +\psi(u_1,v_1,\tau_1)-\psi(u_1',v_1',\tau'_1)]\nonumber\hspace{2mm}\\
+\textstyle\frac{1}{2}\cos[\w_0(\tau_1+\tau'_1) +\psi(u_1,v_1,\tau_1)+\psi(u_1',v_1',\tau'_1)].\hspace{1mm}
\eea
The last cosine term in (\ref{71}) oscillates at approximately the rate $\cos{2\w_0t}$ and the first cosine term has a minimum period of about $2\pi/\Delta\w_0$.  Thus, this last cosine term contributes negligibly to the time average and (\ref{68}) reduces to\footnote{The summation on the right-hand side of (\ref{73}) can be expressed as 
\bea{73'}
{\rm Re}\Big\{\!\!\!\sum\limits_{(u_1,v_1)}\sum\limits_{(u'_1,v'_1)}[b(u_1,v_1,\tau_1)e^{-i\w_0\tau_1}][b(u'_1,v'_1,\tau'_1)e^{-i\w_0\tau_1'}]^*\Big\}\nonumber\\ = {\rm Re}\Big[\big|\!\!\!\sum\limits_{(u,v)}b(u,v,\tau)e^{-i\w_0\tau}\big|^2\Big] = \big|\!\!\!\sum\limits_{(u,v)}b(u,v,\tau)e^{-i\w_0\tau}\big|^2.\nonumber
\eea
Thus, it checks that the right-hand side of (\ref{73}) is a positive quantity.}
\bea{73}
\overline{E^2(r,\pi/2,\phi_1,t)} = \frac{2}{r^2}\sum\limits_{(u_1v_1)}\sum\limits_{(u_1'v_1')}|b(u_1,v_1,\tau_1)b(u_1',v_1',\tau'_1)|\nonumber\hspace{-8mm}\\\cdot \cos[\w_0(\tau_1-\tau'_1) +\psi(u_1,v_1,\tau_1)-\psi(u_1',v_1',\tau'_1)]\hspace{5mm}
\eea
where $\tau_1-\tau_1'$ does not depend on $t$ or $r$.
Similarly, for the angle $\phi_2$ we have
\bea{74}
\overline{E^2(r,\pi/2,\phi_2,t)} = \frac{2}{r^2}\sum\limits_{(u_2v_2)}\sum\limits_{(u_2'v_2')}|b(u_2,v_2,\tau_2)b(u_2',v_2',\tau'_2)|\nonumber\hspace{-8mm}\\\cdot \cos[\w_0(\tau_2-\tau'_2) +\psi(u_2,v_2,\tau_2)-\psi(u_2',v_2',\tau'_2)]\hspace{5mm}
\eea
with
\be{75}
\tau_2 = t-(r +u_2a\phi_2)/c,\;\;\tau'_2 = t-(r +u_2'a\phi_2)/c
\ee
where $\tau_2-\tau_2'$ does not depend on $t$ or $r$.
\par
If the output in (\ref{73}) of the photoelectric detector at $\phi_1$ is correlated with the output in (\ref{74}) of the photoelectric detector at $\phi_2 = \phi_1 + \Delta\phi$ over a long-time average $T\gg 2\pi/\Delta\w_0 \gg T_d$ (see (\ref{P6})), we obtain the second-order correlation function
\be{76}
G^{(2)}(\Delta\phi) = \left\langle\overline{E^2(r,\pi/2,\phi_1,t)}\;\;\overline{E^2(r,\pi/2,\phi_2,t)}\right\rangle.
\ee
Substitution from (\ref{73}) and (\ref{74}) into (\ref{76}) yields
\bea{77}
G^{(2)}(\Delta\phi) = \frac{4}{r^4}\sum\limits_{(uv)_1}\sum\limits_{(uv)'_1}\sum\limits_{(uv)_2}\sum\limits_{(uv)'_2} 
\langle|b_1 b_1'b_2 b_2'|\nonumber\hspace{5mm}\\\cdot\cos[\w_0(\tau_1-\tau'_1) +\psi(u_1,v_1,\tau_1)-\psi(u_1',v_1',\tau'_1)]\nonumber\hspace{1mm}\\ \cdot\cos[\w_0(\tau_2-\tau'_2) +\psi(u_2,v_2,\tau_2)-\psi(u_2',v_2',\tau'_2)]\rangle
\eea
with $b_1 = b(u_1,v_1,\tau_1)$ etc.   For $(uv)_1 = (uv)'_1$, the long-time average approaches zero unless $(uv)_2 = (uv)'_2$ because the phase $[\psi(u_2,v_2,\tau_2)-\psi(u_2',v_2',\tau'_2)]$ continuously varies with time between $0$ and $2\pi$ except for $(uv)_2 = (uv)'_2$.  The contribution to the quadruple summation in (\ref{77}) from $[(uv)_1 = (uv)'_1, (uv)_2 = (uv)'_2]$ is simply
\be{78}
\sum\limits_{(uv)_1}\sum\limits_{(uv)_2}\langle|b_1|^2|b_2|^2\rangle. 
\ee
From (\ref{50}) we can express $|b_1|^2$ and $|b_2|^2$ as
\be{78'}
|b_1|^2 = \frac{(2\pi)^4 F_0^2}{k_0^4(k_0a)^2} + z_1(t),\;\;\;|b_2|^2 = \frac{(2\pi)^4 F_0^2}{k_0^4(k_0a)^2} + z_2(t)
\ee
where $z_1(t)$ and $z_2(t)$ are statistically independent functions whose time averages are zero.  Inserting (\ref{78'}) into (\ref{78}) shows that the $z(t)$'s contribute to the summation only when they are the same, that is,  for $(uv)_1 =(uv)_2$. Thus, they contribute only an amount $O(k_0a)$ and (\ref{78}) evaluates to
\be{78''}
\sum\limits_{(uv)_1}\sum\limits_{(uv)_2}\langle|b_1|^2|b_2|^2\rangle = \frac{(2\pi)^4 F_0^2}{k_0^4(k_0a)^2}\left[\pi(k_0a)^2 + O(k_0a)\right] \approx b_0^4
\ee
with $b_0^2$ defined in (\ref{50sum2}).
\par
Utilization of (\ref{78''}) allows (\ref{77}) to be re-expressed as
\bea{79}
G^{(2)}(\Delta\phi) = \frac{4}{r^4}b_0^4 + \frac{4}{r^4}\hspace{-6mm}\sum\limits_{(uv)_1}^{\hspace{5mm}(uv)_1'\neq(uv)_1}\hspace{-6mm}\sum\limits_{(uv)'_1}\hspace{-6mm}\sum\limits_{(uv)_2}^{\hspace{5mm}(uv)_2'\neq(uv)_2}\hspace{-6mm}\sum\limits_{(uv)'_2} \langle|b_1 b_1'b_2 b_2'|\nonumber\hspace{-6mm}\\\cdot\cos[\w_0(\tau_1-\tau'_1) +\psi(u_1,v_1,\tau_1)-\psi(u_1',v_1',\tau'_1)]\hspace{5mm}\nonumber\\ \cdot\cos[\w_0(\tau_2-\tau'_2) +\psi(u_2,v_2,\tau_2)-\psi(u_2',v_2',\tau'_2)]\rangle.\hspace{2mm}
\eea
The product of the cosines in (\ref{79}) can be rewritten as
\bea{80}
&&\hspace{-5mm}\textstyle\frac{1}{2}\cos[\w_0(\tau_1-\tau'_1+\tau_2-\tau'_2) +\psi_1-\psi_1'+\psi_2-\psi_2']\hspace{-5mm}\\
&&\hspace{-4.5mm}+\textstyle\frac{1}{2}\cos[\w_0(\tau_1-\tau'_1-\tau_2+\tau'_2) +\psi_1-\psi_1'-\psi_2+\psi_2']\hspace{-5mm}\nonumber\\
&&\hspace{-5mm}=\!\!\textstyle\frac{1}{2}\cos\{k_0a[\phi_1(u_1\!-\!u_1')\!+\!\phi_2(u_2\!-\!u_2')] +\psi_1'\!-\!\psi_1\!+\!\psi_2'\!-\!\psi_2\}\nonumber\hspace{-5mm}\\
&&\hspace{-4mm}+\textstyle\frac{1}{2}\cos\{k_0a[\phi_1(u_1\!-\!u_1')\!-\!\phi_2(u_2\!-\!u_2')] \!+\!\psi_1'\!-\!\psi_1\!-\!\psi_2'\!+\!\psi_2\}\nonumber
\eea
with $\psi_1 = \psi(u_1,v_1,\tau_1)$, etc.
The long-time average in (\ref{79}) of the next to the last cosine function in (\ref{80}) approaches zero unless $(uv)_2=(uv)_1'$ and $(uv)'_2=(uv)_1$, in which case this cosine function equals\\[-3mm]
\be{81}
\textstyle\frac{1}{2}\cos[k_0a\Delta\phi(u_1'-u_1) +\psi_{12}'-\psi_{12}]\\[-2mm]
\ee
where
\begin{subequations}
\lbl{82}
\bea{82a}
\psi_{12} = \psi(u_1,v_1,t-r/c -u_1a\phi_1/c)\nonumber\hspace{15mm}\\ - \psi(u_1,v_1,t-r/c -u_1a\phi_2/c)\nonumber\\[-2mm]
\approx \frac{\partial}{\partial t}\psi(u_1,v_1,t-r/c -u_1a\phi_1/c)u_1 a\Delta\phi/c
\eea
\bea{82b}
\mbox{}\nonumber\\[-7mm]
\psi'_{12} = \psi(u'_1,v'_1,t-r/c -u'_1a\phi_1/c)\nonumber\hspace{15mm}\\ - \psi(u'_1,v'_1,t-r/c -u'_1a\phi_2/c)\nonumber\\[-2mm]
\approx \frac{\partial}{\partial t}\psi(u'_1,v'_1,t-r/c -u'_1a\phi_1/c)u'_1 a\Delta\phi/c.
\eea
\end{subequations}
Since $\partial\psi/\partial t$ is on the order of $\Delta\w_0\psi$, (\ref{82}) can be rewritten\\[-3mm]
\begin{subequations}
\lbl{83}
\be{83a}
\psi_{12} = O(\Delta\w_0/\w_0) k_0a\Delta\phi u_1\psi
\ee
\mbox{}\\[-10mm]
\be{83b}
\psi'_{12} =  O(\Delta\w_0/\w_0) k_0a\Delta\phi u'_1\psi
\ee
\end{subequations}
which are negligible when inserted into (\ref{81}) because $\Delta\w_0/\w_0\ll 1$ and we are interested in values of $k_0a\Delta\phi$ on the order of unity.  Thus, (\ref{81}) reduces to simply
\be{84}
\textstyle\frac{1}{2}\cos[k_0a\Delta\phi(u_1'-u_1)].
\ee
By a similar argument, the last cosine function in (\ref{80}) also reduces to (\ref{84}).  In all then, the quadruple summation in (\ref{79}) reduces to a double summation and $G^{(2)}(\Delta\phi)$ becomes\\[-2mm]
\be{85}
\frac{4}{r^4}\bigg(b_0^4 + \hspace{-9mm}\sum\limits_{(uv)_1}^{\hspace{8mm}(uv)_1'\neq(uv)_1}\hspace{-8mm}\sum\limits_{(uv)'_1} \langle|b_1 b_1'|^2\rangle\cos[k_0a\Delta\phi(u_1'-u_1)]\bigg)\\[-1mm]
\ee
where we have applied the corresponding argument to the $b$'s to prove that $|b_2'|\approx |b_1|$ and  $|b_2|\approx |b'_1|$ for $\Delta\w_0/\w_0\ll 1$ and  $k_0a\Delta\phi$ on the order of unity.  Writing $\langle|b_1 b_1'|^2\rangle = \langle|b_1|^2|b_1'|^2\rangle$ and noting from (\ref{50}) and (\ref{50sum2}) that\\[-2mm]
\be{86}
|b_1|^2|b_1'|^2 = \left[\frac{b_0^2}{\pi(k_0a)^2} + z_1(t)\right]\left[\frac{b_0^2}{\pi(k_0a)^2} + z'_1(t)\right]\\[-.5mm]
\ee
where $z_1(t)$ and $z'_1(t)$ are functions of time that time-average to zero, we find\\[-3mm]
\be{87}
\langle|b_1 b_1'|^2\rangle = \langle|b_1|^2|b_1'|^2\rangle = \frac{b_0^4}{\pi^2(k_0a)^4}\\[-1mm] 
\ee
since $(uv)_1'\neq(uv)_1$ and thus $z_1(t) \neq z'_1(t)$ so that $\langle z_1(t) z_1'(t)\rangle = 0$.
The result in (\ref{87}) simplifies (\ref{85}) to\\[-4mm]
\bea{88}
G^{(2)}(\Delta\phi) = \frac{4}{r^4}b_0^4\bigg( 1\hspace{45mm}\nonumber\\ + \frac{1}{\pi^2(k_0a)^4}\hspace{-6mm}\sum\limits_{(uv)_1}^{\hspace{8mm}(uv)_1'\neq(uv)_1}\hspace{-8mm}\sum\limits_{(uv)'_1} \cos[k_0a\Delta\phi(u_1'-u_1)]\bigg).
\eea
\mbox{}\\[-7mm]
\par
Approximating the two-fold double summation by a two-fold double integral converts (\ref{88}) to\\[-4mm]
\bea{89}
G^{(2)}(\Delta\phi) = \frac{4}{r^4}b_0^4\bigg( 1\hspace{48mm}\nonumber\\ + \frac{1}{\pi^2}\int\limits_{-1}^{+1}\int\limits_{-1}^{+1} \int\limits_{-\sqrt{1-u^2}}^{+\sqrt{1-u^2}}\hspace{-5mm}dv \int\limits_{-\sqrt{1-u'^2}}^{+\sqrt{1-u'^2}}\hspace{-5mm}dv'\cos[k_0a\Delta\phi(u'-u)]dudu'\bigg)
\eea
\mbox{}\\[-5mm]
or\\[-4mm]
\bea{90}
&&\hspace{-17mm}G^{(2)}(\Delta\phi) = \frac{4}{r^4}b_0^4\bigg( 1\\ &&\hspace{-18mm}+ \frac{4}{\pi^2}\int\limits_{-1}^{+1}\int\limits_{-1}^{+1} \sqrt{1-u'^2}\sqrt{1-u^2}\, \cos[k_0a\Delta\phi(u'-u)]dudu'\bigg).\hspace{-10mm}\nonumber
\eea
\mbox{}\\[-3mm]
The double integral in (\ref{90}) can be evaluated with the aid of the identity\\[-5mm]
\bea{91}
\cos[k_0a\Delta\phi(u'-u)] = \cos(k_0a\Delta\phi u')\cos(k_0a\Delta\phi u)\nonumber\\ + \sin(k_0a\Delta\phi u')\sin(k_0a\Delta\phi u)
\eea
\mbox{}\\[-7mm]
to get\\[-3mm]
\be{92}
G^{(2)}(\Delta\phi) = \frac{4}{r^4}b_0^4\bigg[ 1 + \frac{4}{\pi^2}\Big(\int\limits_{-1}^{+1} \sqrt{1-u^2} \cos(k_0a\Delta\phi u)du\Big)^2\bigg].
\ee
As in (\ref{53}), the integral evaluates to a first-order Bessel function to yield the second-order correlation in the form
\bea{93}
G^{(2)}(\Delta\phi) = \frac{4}{r^4}b_0^4\left[ 1 + \left(\frac{2J_1(k_0a\Delta\phi)}{k_0a\Delta\phi}\right)^2\right]
\eea
which can be expressed in terms of the normalized first-order correlation function in (\ref{55}) as\\[-3mm]
\be{94}
G^{(2)}(\Delta\phi) = \frac{4}{r^4}b_0^4\left( 1 + \left[g^{(1)}(\Delta\phi)\right]^2\right).\\[-1mm]
\ee
Defining the normalized second-order correlation function as\\[-3mm]
\be{95}
g^{(2)}(\Delta\phi) = \frac{G^{(2)}(\Delta\phi)}{\left\langle\overline{E^2(r,\pi/2,\phi,t)}\right\rangle^2}\\[-1mm]
\ee
and noting from (\ref{29}), (\ref{54}), and (\ref{50sum2}) that\\[-3mm]
\be{96}
\left\langle\overline{E^2(r,\pi/2,\phi,t)}\right\rangle^2 = \left[G^{(1)}(0)\right]^2 = \frac{4}{r^4}b_0^4\\[-1mm]
\ee
one finds the following simple relationship between the normalized first- and second-order correlation functions
\be{97}
g^{(2)}(\Delta\phi) =1 + \left[\frac{2J_1(k_0a\Delta\phi)}{k_0a\Delta\phi}\right]^2 =  1 + \left[g^{(1)}(\Delta\phi)\right]^2.
\ee
Hanbury Brown--Twiss intensity stellar interferometry uses this result to find the normalized first-order correlation  $g^{(1)}(\Delta\phi)$ from a measurement of the normalized second-order correlation $g^{(2)}(\Delta\phi)$, which requires only the measurement of radiation intensities by two photoelectric detectors.
\section{Conclusion}\lbl{sec:Conclusion}
Using a realistic model of the sun and other stars at optical wavelengths as a spherical antenna composed of statistically independent volume sources, rather than the usual planar aperture-field, circular-disc model of the sun and other stars, a self-contained, straightforward, detailed derivation is given for the narrow-bandwidth received fields and speckle patterns radiated by the star and for the first- and second-order correlation functions satisfied by the fields and measured in Michelson phase stellar interferometry and Hanbury Brown--Twiss intensity stellar interferometry, respectively.  The derivation hinges on the use of a newly derived spherical wave expansion that involves a Fourier series in both the spherical angles $\theta$ and $\phi$ and on determining the time averages of the associated spherical-wave coefficients as required by the assumed Lambertian radiation of the sun and other stars within the visible spectrum.  It is shown that the $ka$ bandlimit that holds for the order of the spherical Hankel functions in the spherical wave expansions of electrically large, nonsuper-reactive, coherent sources also applies to the incoherent stellar sources, and that the $\pi(k_0a)^2$ quasi-monochromatic spherical mode coefficients are uncorrelated (their temporal cross correlation is zero).  
\par
Working directly with the real-valued time-domain fields and their correlations, and without having to invoke van Cittert-Zernicke, central limit, or moment theorems, the expression for the normalized first-order correlation function $g^{(1)}(\Delta\phi)$ in (\ref{55}), used in Michelson phase stellar interferometer measurements, and the expression for the normalized second-order correlation function $g^{(2)}(\Delta\phi)$ in (\ref{97}), used in Hanbury Brown--Twiss intensity stellar interferometer measurements, are derived and shown to satisfy the simple relationship $g^{(2)}(\Delta\phi) =  1 + \left[g^{(1)}(\Delta\phi)\right]^2$.  For Lambertian radiation (no intensity  taper), the classic angular separation distance of $\Delta\phi_0 = 1.22\lambda_0/D$ is found for the first null of both the temporally averaged and spatially averaged first-order correlation functions, where $\lambda_0$ is the mean wavelength of the assumed narrow bandwidth of the measured radiation and $D=2a$ is the diameter of the star.  Although the stars are as\-sumed to be Lambertian radiators throughout the analysis of this paper, if it is necessary to account for limb effects, this can be done by inserting the observed intensity taper of the star into the intensity parameter, as explained in Section \ref{subsec:Lambertian}.  
\par
Among the advantages of working with real-valued fields directly in the time domain are that explicit expressions are found for the quasi-monochromatic wave-packet fields radiated by the spherical star and that new criteria, much less restrictive than the Rayleigh distance for coherent sources, are revealed for the minimum distance at which the large-argument approximation for the Hankel function can be used to determine the radiated fields from the incoherent stellar sources.   The terminology, concepts, and methodology used in the direct real-valued time-domain solution for the fields and correlation functions of spherical stars are first introduced by similarly solving the much simpler problem of a linear array of randomly excited scalar-field (acoustic) point sources.
\section*{Acknowledgment}
The paper benefited from the thoughtful comments and suggestions of Professor Francesco Monticone.  This work was supported in part under the U.S. Air Force Office of Scientific Research Contract \# FA9550-19-1-0097 through Dr. Arje Nachman.
\end{document}